\documentclass[11pt]{amsart}
\usepackage{graphicx}

\newtheorem{thm}{Theorem}[section]

\newtheorem{lem}[thm]{Lemma}
\newtheorem{prop}[thm]{Proposition}
\newtheorem{problem}[thm]{Problem}
\theoremstyle{definition}
\newtheorem{defn}[thm]{Definition}
\theoremstyle{remark}
\newtheorem{rem}[thm]{Remark}
\newtheorem{ex}[thm]{Example}
\numberwithin{equation}{section}

\newcommand{\abs}[1]{\left\vert#1\right\vert}
\newcommand{\set}[1]{\left\{#1\right\}}
\newcommand{\Real}{\mathbb R}
\newcommand{\Natural}{\mathbb N}

\newcommand{\B}{\mathcal{B}}

\newcommand{\Pre}{\mathcal{P}}
\newcommand{\such}{\ | \ }
\newcommand{\ud}{\mathrm d}

\newcommand{\prob}{\mathbb{P}}
\newcommand{\qprob}{\mathbb{Q}}
\newcommand{\expec}{\mathbb{E}}

\newcommand{\basis}{(\Omega, \mathcal{F}, \mathbf{F}, \prob)}
\newcommand{\filtration}{\mathbf{F} = \pare{\mathcal{F}_t}_{t \in \Real_+}}
\newcommand{\F}{\mathcal{F}}
\newcommand{\G}{\mathcal{G}}

\newcommand{\p}{\mathrm{p}}

\newcommand{\dist}{\mathrm{dist}}
\newcommand{\g}{\mathfrak{g}}
\newcommand{\rel}{\mathfrak{rel}}
\newcommand{\R}{\mathcal{R}}
\newcommand{\Z}{\mathcal{Z}}

\newcommand{\pare}[1]{\left(#1\right)}
\newcommand{\bra}[1]{\left[#1\right]}
\newcommand{\dbra}[1]{[\kern-0.15em[ #1 ]\kern-0.15em]}
\newcommand{\N}{\mathfrak{N}}
\newcommand{\Af}{\mathfrak{A}}

\newcommand{\Uc}{\mathcal{U}}

\newcommand{\K}{\mathfrak{C}}
\newcommand{\Kc}{\check{\mathfrak{C}}}

\newcommand{\I}{\mathfrak{I}}
\newcommand{\Log}{\mathcal L}
\newcommand{\Exp}{\mathcal E}
\newcommand{\indic}{\mathbb{I}}
\newcommand{\Var}{\mathrm{Var}}
\newcommand{\co}{\mathsf{c}}
\newcommand{\loc}{\mathsf{loc}}

\newcommand{\Def}{\mathfrak{D}}
\newcommand{\cone}{\mathsf{cone}}
\newcommand{\inner}[2]{\left \langle #1 , #2 \right \rangle}
\newcommand{\num}{num\'eraire }
\newcommand{\hx}{x \indic_{\{ |x| \leq 1 \}}}
\newcommand{\hbarx}{x \indic_{\{ |x| > 1 \}}}
\newcommand{\Omegat}{\tilde{\Omega}}
\newcommand{\omegat}{\tilde{\omega}}

\newcommand{\NUIP}{\emph{NUIP }}

\newcommand{\NUPBR}{\emph{NUPBR }}
\newcommand{\NA}{\emph{NA }}

\newcommand{\NFLVRC}{NFLVR$_\K$ }
\newcommand{\NUPBRC}{NUPBR$_\K$ }
\newcommand{\NAC}{NA$_\K$ }
\newcommand{\ESMMC}{ESMM$_\K$ }
\newcommand{\ELMMC}{ELMM$_\K$ }

\newcommand{\QprobC}{\mathfrak{M}_\K}

\begin{document}

\title
{The Num\'eraire Portfolio in Semimartingale Financial Models}%
\author[I. Karatzas]{Ioannis Karatzas}
\address{Mathematics and Statistics Departments, Columbia University, NY 10027}
\email{ik@math.columbia.edu}
\author[C. Kardaras]{Constantinos Kardaras}
\address{Mathematics and Statistics Department, Boston University, MA 02215}
\email{kardaras@bu.edu}

\thanks{Results in this paper are drawn in part from the second author's doctoral
 dissertation \cite{K: thesis}.
Work   partially supported by the National Science Foundation, under
grant NSF-DMS-06-01774.}

\date{\today}%
\begin{abstract}
We study the existence of the \textsl{\num portfolio} under
predictable convex constraints in a general semimartingale model of
a financial market. The \num portfolio generates a wealth process,
with respect to which the relative wealth processes of all other
portfolios are supermartingales. Necessary and sufficient conditions
for the existence of the \num portfolio are obtained in terms of the
triplet of predictable characteristics of the asset price process.
This characterization is then used to obtain further necessary and
sufficient conditions, in terms of a no-free-lunch-type notion. In
particular, the full strength of the ``No Free Lunch with Vanishing
Risk'' (NFLVR) is not needed, only the weaker ``No Unbounded Profit
with Bounded Risk'' (NUPBR) condition that involves the boundedness
in probability of the terminal values of wealth processes. We show
that this notion is the minimal a-priori assumption required in
order to proceed with utility optimization. The fact that it is
expressed entirely in terms of predictable characteristics makes it
easy to check, something that the stronger NFLVR condition lacks.
\end{abstract}

\maketitle

\setcounter{section}{-1}

\section{Introduction} \label{sec: Introduction and some Notation}

\subsection{Background and Discussion of Results}

A broad class of models, that have been used extensively in
Stochastic Finance, are those for which the price processes of
financial instruments are considered to evolve as semimartingales.
The concept of a semimartingale is very intuitive: it connotes a
process that can be decomposed into a \emph{finite variation} term
that represents the ``signal'', and a \emph{local martingale} term
that represents the ``noise''. The reasons for the ubiquitousness of
semimartingales in modeling financial asset prices are by now pretty
well-understood --- see for example Delbaen and Schachermayer
\cite{DS: FTAP locally bdd}, where it is shown that restricting
ourselves to the realm of locally bounded stock prices, and agreeing
that we should banish arbitrage by use of simple ``buy-and-hold''
strategies, the price process \emph{has} to be a semimartingale.
Discrete-time models can be embedded in this class, as can processes
with independent increments and many other Markov processes, such as
solutions to stochastic differential equations. Models that are not
encompassed, but have received attention, include price-processes
driven by fractional Brownian motion.

\smallskip

In this paper we consider a general semimartingale model and make no
further mathematical assumptions. On the economic side, we assume
that assets have their prices determined exogenously, and can be
traded without ``frictions'': transaction costs are non-existent or
negligible. Our main concern will be a problem of dynamic stochastic
optimization: to find a trading strategy whose wealth appears
``better'' when compared to the wealth generated by any other
strategy, in the sense that the ratio of the two processes is a
supermartingale. If such a strategy exists, it is essentially
unique and it is called \textsl{\num portfolio}. Necessary and
sufficient conditions for the \num portfolio to exist are derived,
in terms of the triplet of \emph{predictable characteristics}
(loosely speaking these are the drift, the volatility co\"efficient,
and the jump intensity) of the stock-price returns.

Sufficient conditions for the existence of the \num portfolio are
established in Goll and Kallsen \cite{Goll - Kallsen: log-optimal},
who focus on the (almost equivalent) problem of maximizing expected
logarithmic utility. These authors show that their conditions are
also necessary, under the following assumptions: the problem of
maximizing the expected log-utility from terminal wealth has a
finite value, no constraints are enforced on strategies, and NFLVR
holds. Becherer \cite{Becherer} also discusses how, under these
assumptions, the \num portfolio exists and coincides with the
log-optimal one. In both these papers, deep results of Kramkov and
Schachermayer \cite{Kramkov-Schachermayer: Asymptotic elast and util
function} on utility maximization are invoked.

Here we follow a bare-hands approach which enables us to obtain
stronger results. First, the assumption of finite expected
log-utility is dropped entirely; there should be no reason for it
anyhow, since we are not working on the problem of log-utility
optimization. Secondly, general closed convex constraints on
portfolio choice can be enforced, as long as they unfold in a
predictable manner. Thirdly, and perhaps most controversially, we
drop the NFLVR assumption: \emph{no} normative assumption   is
imposed on the model. It turns out that the \num portfolio can exist
even when the classical \emph{No Arbitrage} (NA) condition fails.

In the context of stochastic portfolio theory, we feel there is no
need for no-free-lunch assumptions to begin with: the r\^ole of
optimization should be to find and utilize arbitrage opportunities
in the market, rather than ban the model. It is actually possible
that the optimal strategy of an investor is \emph{not} an arbitrage
(an example involves the notorious three-dimensional Bessel process and
can be found in \S 3.3.3 of the present paper). The usual practice
of assuming that we can invest unconditionally on arbitrages breaks
down because of credit limit constraints: arbitrages are sure to
generate, at a fixed future date,  more capital than initially
invested; but they can do pretty badly in the meantime, and this imposes
an upper bound on the capital that can be invested. There exists an
even more severe problem when trying to argue that arbitrages should
be banned: in very general semimartingale financial markets
\emph{there does not seem to exist any computationally feasible way
of deciding whether arbitrages exist or not}. This goes hand-in-hand
with the fact that existence of equivalent martingale measures
--- its remarkable theoretical importance notwithstanding --- is
\emph{not easy to check}, at least by looking directly at the
dynamics of the stock-price process.

Our second main result comes hopefully to shed some light on this
situation. Having made no model assumptions when initially
trying to decide whether the \num portfolio exists, we now take a
step backwards and in the opposite-than-usual direction: we ask
ourselves \emph{what the existence of the \num portfolio can tell us
about free-lunch-like opportunities in the market}. Here, the
necessary and sufficient condition for existence of the \num
portfolio is the boundedness in probability of the collection of
terminal wealths attainable by trading (``no unbounded profit with
bounded risk'', NUPBR  for short). This is one of the two conditions
that comprise NFLVR; what remains of course is the NA condition. In
the spirit of the Fundamental Theorem of Asset Pricing, we show that
another mathematical equivalence to the NUPBR condition is
\emph{existence of equivalent supermartingale deflators}, a concept
closely related but strictly weaker than Equivalent Martingale
Measures. A similar result appears in Christensen and Larsen
\cite{Chr-Lar: NA and GOP}, where the results of Kramkov and
Schachermayer \cite{Kramkov-Schachermayer: Asymptotic elast and util
function} are again used.

We then go on further, and ask how severe this NUPBR assumption
really is. The answer is simple: when this condition fails, one
cannot do utility optimization for \emph{any} utility function;
conversely if this assumption holds, one can proceed with utility
maximization as usual. The main advantage of not assuming the full
NFLVR condition is that, \emph{there is a direct way of checking
the validity of the weaker \NUPBR condition in terms of the
predictable characteristics of the price process}. No such
characterization exists for the NA condition, as Example 3.7 in
subsection \ref{subsec: beyond FTAP} demonstrates. Furthermore,
our result can be used to understand the gap between the concepts
of NA and the stronger NFLVR; \emph{the existence of the \num
portfolio is exactly the bridge needed to take us from \NA to}
NFLVR. This was known for continuous-path processes since the
paper \cite{DS: FTAP absolutely continuous} of Delbaen and
Schachermayer; here we do it for the general case.

\subsection{Synopsis}

After this short subsection, in the remainder of this section we
recall probabilistic concepts to be used throughout.

Section \ref{sec: The Market, Investments and Constraints}
introduces the financial market model, the ways in which agents can
invest in this market, and the constraints that are faced. In
section \ref{sec: Numeraire portfolio} we introduce the \num
portfolio. We discuss how it relates to other notions, and conclude
with our main Theorem \ref{thm: existence of the num for semimarts} that provides necessary and sufficient conditions for the existence
of the \num portfolio in terms of the predictable characteristics of
the stock-price processes. Section \ref{sec: Arb, ESMM and supermart
defl} deals with the connections between the \num portfolio and free
lunches. The main result there is Theorem \ref{thm: num iff
def-non-empty iff NUPBR}, which can be seen as another version of
the Fundamental Theorem of Asset Pricing.

Certain proofs that are not given in sections \ref{sec: Numeraire
portfolio} and \ref{sec: Arb, ESMM and supermart defl} occupy the
next four sections. In the self-contained section \ref{sec: The NUIP
condition} we describe necessary and sufficient conditions for the
existence of wealth processes that are increasing and not constant.
In section \ref{sec: The Numeraire for General Semimartingales} we
prove our main Theorem \ref{thm: existence of the num for
semimarts}. Section \ref{sec: Rates of Conv to Zero of Pos
Supermarts} contains a result on rates of convergence to zero of
positive supermartingales, which is used to study an asymptotic
optimality property of the \num portfolio. Finally, section
\ref{sec: proof that with num you can never go wrong} completes
proving our second main Theorem \ref{thm: num iff def-non-empty iff
NUPBR}.

In order to stay as self-contained as possible, Appendices are
included on: (A) measurable random subsets and selections; (B)
semimartingales up to infinity and the corresponding ``stochastic
integration up to infinity''; and (C) $\sigma$-localization.

\subsection{Remarks of probabilistic nature}

For results concerning the general theory of stochastic processes
described below, we refer the reader to the book \cite{Jacod -
Shiryaev} of Jacod and Shiryaev,  especially the first two
chapters.

We are given a  stochastic basis  $\basis$, where the filtration
$\filtration$ is assumed to satisfy the usual hypotheses  of
right-continuity and augmentation by the $\prob$-null sets. The
probability measure  $\prob$ will be fixed throughout and every
formula, relationship, etc. is supposed to be valid $\prob$-almost
surely ($\prob$-a.s.)

The \textsl{predictable $\sigma$-algebra} on the \textsl{base space}
$\Omega \times \Real_+$ will be denoted by $\Pre$ --- if $\pi$ is a
$d$-dimensional predictable process we write $\pi \in
\Pre(\Real^d)$. For any adapted, right-continuous process $Y$ that
admits left-hand limits, we denote by $Y_-$ its predictable
\textsl{left-continuous version} and its \textsl{jump process} is
$\Delta Y := Y - Y_-$.

For a $d$-dimensional semimartingale $X$ and $\pi \in
\Pre(\Real^d)$, we denote by $\pi \cdot X$ the \textsl{stochastic
integral} process, whenever this makes sense, in which case we shall
be referring to $\pi$ as being \textsl{$X$-integrable}. We are
assuming \textsl{vector stochastic integration}, good accounts of
which can be found in \cite{Bichteler: stoch. integration},
\cite{Cherny-Shiryaev: Vector stoch. integral} and \cite{Jacod -
Shiryaev}. For two semimartingales $X$ and $Y$, $[X,Y] := X Y - X_-
\cdot Y - Y_- \cdot X$ is their \textsl{quadratic covariation}
process.

The \textsl{stochastic exponential} $\Exp(Y)$ of the scalar
semimartingale $Y$ is the unique solution $Z$ of the stochastic
integral equation $Z = 1 + Z_- \cdot Y$ and is given by
\begin{equation} \label{eq: stoch expo}
\Exp(Y) = \exp \set{Y - \frac{1}{2} [Y^\co, Y^\co]}  \cdot
 \prod_{s \leq \cdot} \set{\pare{1 + \Delta Y_s} \exp(-\Delta
Y_s)},
\end{equation}
where $Y^\co$ denotes the \textsl{continuous martingale part} of the
semimartingale $Y$. The stochastic exponential $Z = \Exp(Y)$
satisfies $Z > 0$ and $Z_- > 0$ if and only if $\Delta Y > -1$.
Given a semimartingale $Z$ which satisfies $Z > 0$ and $Z_- > 0$, we
can invert the stochastic exponential operator and get the
\textsl{stochastic logarithm} $\Log(Z)$, which is defined as
$\Log(Z) := (1/Z_-) \cdot Z$ and satisfies $\Delta \Log(Z) > -1$.

\section{The Market, Investments, and Constraints} \label{sec: The Market, Investments and
Constraints}

\subsection{The asset-prices model} \label{subsec: stock-price model}

On the given stochastic basis $(\Omega, \F, \mathbf{F}, \prob)$ we
consider $d$ strictly positive semimartingales $S^1, \ldots, S^d$
that model the prices of $d$ assets; we shall refer to these as
\textsl{stocks}. There is also another process $S^0$, representing
the \textsl{money market}  or \textsl{bank account} --- this asset
is a ``benchmark'', in the sense that wealth processes will be
quoted in units of $S^0$ and not nominally. As is usually done in
this field, we assume $S^0 \equiv 1$, making $S^1, \ldots, S^d$
already \emph{discounted} asset prices. This does not affect the
generality of the discussion, since otherwise we can divide all
$S^i$, $i = 0, 1, \ldots, d$ by $S^0$.

For all $i = 1, \ldots, d$, $S^i$ and $S_-^i$ are strictly positive;
therefore, there exists a $d$-dimensional semimartingale $X \equiv
(X^1, \ldots, X^d)$ with $X_0 = 0$, $\Delta X^i > -1$ and $S^i =
S^i_0 \,\Exp(X^i)$ for $i=1,\ldots, d$. We interpret $X$ as the
\emph{discounted returns} that generate the asset prices $S$ in a
multiplicative way. In our discussion we shall be using the returns
process $X$, not the stock-price process $S$ directly.

Our financial planning horizon will be $\dbra{0, T} := \{(\omega,
t) \in \Omega \times \Real_+ \such t \leq T (\omega) \}$ where $T$
is a \emph{possibly infinite-valued stopping time}. Observe that,
as usual, even if $T$ takes infinite values, the time-point at
infinity is \emph{not} included in the definition of $\dbra{0,
T}$. All processes then will be considered as being constant and
equal to their value at $T$ for all times after $T$, i.e., every
process $Z$ is equal to the \textsl{stopped process} at time $T$,
is defined via $Z_t^T := Z_{t \wedge T}$ for all $t \in \Real_+$.
We can assume further, without loss of generality, that $\F_0$ is
$\prob$-trivial (thus all $\F_0$-measurable random variables are
constants) and that $\F = \F_{T} := \bigvee_{t \in \Real_+} \F_{t
\wedge T}$.

\begin{rem}
Under our model we have $S^i > 0$ and $S^i_- > 0$; to be in par with
the papers \cite{DS: FTAP locally bdd, DS: FTAP unbounded} on
no-free-lunch criteria, we should allow for models with possibly
negative asset prices (for example, forward contracts). All our
subsequent work carries to these models vis-a-vis. We choose to work
in the above set-up because it is somehow more intuitive and
applicable: almost every model used in practice is written in this
way. A follow-up to this discussion is subsection \ref{subsec: on
additive model}.
\end{rem}

The predictable characteristics of the returns process $X$ will be
 very important in our discussion. To this end, we fix the
\textsl{canonical truncation function} $x \mapsto \hx$ (we use
$\indic_A$ to denote the indicator function of some set
$A$) and write the \textsl{canonical decomposition} of the
semimartingale $X$, namely:
\begin{equation} \label{eq: canonical representation}
X = X^\co + B + \bra{\hx} * (\mu - \eta) + \bra{\hbarx} * \mu\,.
\end{equation}
Some remarks are in order. Here, $\mu$ is the \textsl{jump measure}
of $X$, i.e., the random counting measure on $\Real_+ \times
\Real^d$ defined by
\begin{equation} \label{eq: jump measure}
\mu ([0,t] \times A) := \sum_{0 \leq s \leq t} \indic_{A \setminus \{0\}} (\Delta X_s), \ \ \text{ for } t \in
\Real_+ \text{ and } A \subseteq \Real^d.
\end{equation}
Thus, the last process in (\ref{eq: canonical representation}) is
just $\, \bra{\hbarx} * \mu \equiv \sum_{0 \leq s \leq \cdot} \Delta
X_s \indic_{\{ | \Delta X_s | > 1\}}\,$, the \emph{sum of the
``big'' jumps of $X$}; throughout the paper, the asterisk denotes
integration with respect to random measures. Once this term is
subtracted from $X$, what remains is a semimartingale with bounded
jumps, thus a \emph{special} semimartingale. This, in turn, can be
decomposed uniquely
 into a \emph{predictable finite variation} part, denoted by $B$ in (\ref{eq: canonical representation}), and a
\emph{local martingale} part. Finally, this last local martingale
part can be decomposed further: into its \textsl{continuous} part,
denoted by $X^\co$ in \eqref{eq: canonical representation}; and its
\textsl{purely discontinuous} part, identified as the local
martingale $\bra{\hx} * (\mu - \eta)$. Here, $\eta$ is the
\textsl{predictable compensator} of the measure $\mu$, so the purely
discontinuous part is just a \emph{compensated sum of   ``small''
jumps}.

We introduce the \textsl{quadratic covariation} process $C :=
[X^\co, X^\co]$ of $X^\co$, call $(B, C, \eta)$ the \textsl{triplet
of predictable characteristics} of $X$, and define the predictable
increasing scalar process $G := \sum_{i=1}^d
\pare{ C^{i,i} + \Var(B^i) + [1 \wedge |x^i|^2] * \eta}$. Then, all three
$B$, $C$, and $\eta$ are absolutely continuous with respect to $G$, thus
\begin{equation} \label{eq: triplet of pred characteristics}
B = b \cdot G, \ C = c \cdot G, \text{ and } \eta = G \otimes
\nu\,.
\end{equation}
Here  $b$, $c$ and $\nu$ are predictable; $b$ is a vector process,
$c$   a nonnegative-definite matrix-valued process, and $\nu$ a
process with values in the set of L\'evy measures; the symbol
``$\otimes$'' denotes  product measure. Note that \emph{any}
$\widetilde{G}$ with $\ud \widetilde{G}_t \sim \ud G_t$ can be
used in place of $G$; the actual choice of increasing process $G$
reflects the notion of \textsl{operational clock} (as opposed to
the natural time flow, described by $t$). In an abuse of
terminology, we shall refer to $(b, c, \nu)$ also as the
\textsl{triplet of predictable characteristics of} $X$; this
depends on $G$, but the validity of all results not.

\begin{rem} \label{rem: on characteristics being nice everywhere}
The properties of $c$ being a symmetric nonnegative-definite process
and $\nu$ a L\'evy-measure-valued process in general hold $\prob
\otimes G$-a.e. We shall assume that they hold \emph{everywhere} on
$\dbra{0, T}$; we can always do this by altering them on a
predictable set of $\prob \otimes G$-measure zero to be $c \equiv 0$
and $\nu \equiv 0$ (see \cite{Jacod - Shiryaev}).
\end{rem}

\begin{rem}
If $X$ is \textsl{quasi-left-continuous} (i.e., if no jumps occur
at predictable times), $G$ is continuous; but if we want to
include discrete-time models in our discussion, we must allow for
$G$ to have jumps. Since $C$ is continuous and (\ref{eq: canonical
representation}) gives $\expec[\Delta X_\tau \indic_{\{ |\Delta
X_\tau| \leq 1\}} \such \F_{\tau-}] = \Delta B_\tau$ for every
predictable time $\tau$, we get
\begin{equation}
    \label{eq: when G jumps}
c = 0~ \text{ and }~ b = \int \hx \nu (\ud x), ~~\text{ on the
predictable set} ~\{\Delta G > 0\}.
\end{equation}
\end{rem}

The following concept of {\sl drift rate} will be used throughout
the paper.

\begin{defn} \label{dfn: drift rate}
Let $X$ be \emph{any} semimartingale with canonical representation
(\ref{eq: canonical representation}), and consider the process $G$
such that (\ref{eq: triplet of pred characteristics}) hold. On
$\big\{ \int |x| \indic_{\{ |x| > 1\}} \nu(\ud x) < \infty \big\}$,
the \textsl{drift rate} (with respect to $G$) of $X$ is defined as
the expression $b + \int x \indic_{\{ |x| > 1\}} \nu(\ud x)$.
\end{defn}

The range of definition $\set{\int |x| \indic_{\{ |x| > 1\}} \nu(\ud
x) < \infty}$ for the drift rate does not depend on the choice of
operational clock $G$,  though the drift rate itself does. Whenever
the increasing process $\bra{ |x| \indic_{\{ |x| > 1\}}} * \eta =
\pare{\int |x| \indic_{\{ |x| > 1\}} \nu(\ud x)} \cdot G$ is
\emph{finite}  (this happens if and only if $X$ is a \emph{special}
semimartingale),   the predictable process $$B + \bra{\hbarx} * \eta
=
\pare{b + \int x \indic_{\{ |x| > 1\}} \nu(\ud x)} \cdot G$$ is called the \textsl{drift process} of
$X$. If drift processes exist, drift rates exist too; the converse
is not true. Semimartingales that are \emph{not} special might have
well-defined drift rates; for instance, a $\sigma$-martingale is   a
semimartingale with  drift rate identically equal to zero. See
Appendix \ref{subsec: sigma localization} on $\sigma$-localization
for further discussion.

\subsection{Portfolios and Wealth processes}

A financial agent starts with some positive initial capital, which
we normalize to $W_0=1$, and can invest in the stocks by choosing a
\textsl{portfolio} represented by a predictable, $d$-dimensional and
$X$-integrable process $\pi$. With $\pi^i_t$ representing the
\emph{proportion of current wealth} invested in stock $i$ at time
$t$, $\pi_t^0 := 1 - \sum_{i=1}^d \pi^i_t$ is the proportion
invested in the money market.

\smallskip
Some restrictions have to be enforced, so that the agent cannot use
so-called \emph{doubling strategies}. The assumption prevailing in
this context is that the wealth process should be uniformly bounded
from below by some constant --- a \emph{credit limit} that the agent
faces. We shall set this credit limit at zero; one can regard this
as just shifting the wealth process to some constant, and working
with this relative credit line instead of the absolute one.

The above discussion leads to the following definition: a wealth
process will be called \textsl{admissible}, if it and its
left-continuous version stay  strictly positive. Let us denote the
\emph{discounted} wealth process generated from a portfolio $\pi$ by
$W^\pi$; we must have $W^\pi > 0$ and $W_-^\pi > 0$, as well as
\begin{equation} \label{eq: wealth process}
\frac{\ud W^\pi_t}{W^\pi_{t-}} = \sum_{i = 0}^d \pi^i_t \, \frac{\ud
S^i_t}{ S^i_{t-}} = \sum_{i = 1}^d \pi^i_t \, \ud X^i_t = \pi^\top_t
\ud X_t, \ \textrm{ equivalently } W^\pi = \Exp(\pi \cdot X).
\end{equation}

\subsection{Further constraints on portfolios} \label{subsec: constraints on strategies}

We start with an example in order to motivate Definition \ref{dfn:
K-constrained strategies} below.

\begin{ex} \label{ex: short-sale constraints}
Suppose that the agent is prevented from selling stock short. This
means $\pi^i \geq 0$ for all $i=1, \ldots, d$, or that $\pi (\omega,
t) \in (\Real_+)^d$ for all $(\omega, t) \in \dbra{0,T}$. If we
further prohibit borrowing from the money market, then also $\pi^0
\geq 0$; setting $\K:= \{ \p \in \Real^d \such \p^i \geq 0 \text{
and } \sum_{i=1}^d \p^i \leq 1 \}$, the prohibition of short sales
and borrowing translates into the requirement $\pi (\omega, t) \in
\K$ for all $(\omega, t) \in \dbra{0,T}$.
\end{ex}

The example leads us to consider all possible constraints that can
arise this way; although in the above particular case the set $\K$ was
non-random, we shall soon encounter situations where the
constraints depend on both time and the path.

\begin{defn} \label{dfn: K-constrained strategies}
Consider a set-valued process $\K : \dbra{0,T} \rightarrow \B
(\Real^d)$, where $\B (\Real^d)$ is the Borel $\sigma$-algebra on
$\Real^d$. A $\pi \in \Pre(\Real^d)$ will be called
\textsl{$\K$-constrained}, if   $\pi (\omega, t) \in $ $\K(\omega,
t)$ for all $(\omega, t) \in \dbra{0,T}$. We denote by $\Pi_\K$ the
class of all $\K$-constrained, predictable and $X$-integrable
processes that satisfy  $\pi^\top \Delta X > -1\,$.
\end{defn}

The  requirement $\pi^\top \Delta X > -1$ is there to ensure that we
can define the admissible wealth process $W^\pi$, i.e., that the
wealth will remain strictly positive. Let us use this requirement to
give other constraints of this type. Since these actually follow
from the definitions, they will not constrain the wealth processes
further; the point is that we can always include them in our
constraint set.

\begin{ex} \label{ex: natural constraints} (\textsc{Natural Constraints}).
An admissible strategy generates a wealth process that starts
positive and stays positive. Thus, if $W^\pi = \Exp(\pi \cdot X)$,
then we have $\Delta W^\pi \geq - W^\pi_-$, or $\pi^\top \Delta X
\geq -1$. Recalling the definition of the random measure $\nu$ from
(\ref{eq: triplet of pred characteristics}), we see that an
equivalent requirement is
\[
\nu [\pi^\top x < - 1] \equiv \nu [\{ x \in \Real^d \such \pi^\top
x < - 1 \}] = 0, ~\ \prob \otimes G \textrm{-almost everywhere}\,;
\]
Define now \textsl{the random set-valued process of natural
constraints}
\begin{equation} \label{eq: natural constraints}
\K_0 := \{ \p \in \Real^d \such \nu \,[\p^\top x < - 1] = 0 \}
\end{equation}
(randomness comes through $\nu$). Since $\pi^\top X > - 1$, $\pi \in
\Pi_\K$ implies $\pi \in \Pi_{\K \cap \K_0}$.
\end{ex}

Note that $\K_0$ is not deterministic in general --- random
constraints are not introduced just for the sake of generality, but
because they arise naturally in portfolio choice settings. In
subsection \ref{subsec: structure of pred convex constraints}, we
shall impose more structure on the set-valued process $\K\,$: {\sl
convexity, closedness} and {\sl predictability.} The above Examples
\ref{ex: short-sale constraints} and \ref{ex: natural constraints}
have these properties; the ``predictability structure'' should be
clear for $\K_0$, which involves the predictable process $\nu$.

\smallskip

\section{The Num\'eraire Portfolio: Definitions, General Discussion, and
Predictable Characterization} \label{sec: Numeraire portfolio}

\subsection{The \num portfolio}

The following is a central notion of the paper.

\begin{defn} \label{dfn: numeraire}
A process $\rho \in \Pi_\K$ will be called \textsl{\num
portfolio}, if for every $\pi \in \Pi_\K$ the \textsl{relative
wealth process}  $\,W^\pi / W^\rho\,$ is a supermartingale.
\end{defn}

The term ``\num portfolio'' was introduced by Long \cite{Long}; he
defined it as a portfolio $\rho$ that makes $W^\pi / W^\rho$ a
\emph{martingale} for every portfolio $\pi$, then went on to show
that this requirement is equivalent, under some additional
assumptions, to absence of arbitrage for discrete-time and
It\^o-process models. Some authors give the \num portfolio other
names as \textsl{growth optimal} and \textsl{benchmark} (see for
example Platen \cite{Platen: benchmark} who uses the
``num\'eraire'' property as an approach to derivatives pricing,
portfolio optimization, etc.). Definition \ref{dfn: numeraire} in
this form first appears in Becherer \cite{Becherer}, where we send
the reader for the history of this concept. An observation from
that paper is that the wealth process generated by \num portfolios
is unique: if there are two \num portfolios $\rho_1$ and $\rho_2$
in $\Pi_\K$, then both $W^{\rho_1} / W^{\rho_2}$ and $W^{\rho_2} /
W^{\rho_1}$ are supermartingales and Jensen's inequality shows
that they are equal.

Observe that $W^\rho_T$ is always well-defined, even on $\{ T =
\infty \}$, since $1/W^\rho$ is a positive supermartingale and the
supermartingale convergence theorem implies that $W^\rho_T$ exists,
thought it might take the value $+ \infty$ on $\{ T = \infty \}$. A
condition of the form $W^\rho_T < + \infty$ will be essential when
we consider free lunches in section \ref{sec: Arb, ESMM and
supermart defl}.

\begin{rem}
The \num portfolio was introduced in Definition \ref{dfn: numeraire}
as the solution to some sort of optimization problem. It has at
least four more optimality properties that we now mention; these
have already been noted in the literature --- check the appropriate
places in the paper where they are further discussed for references.
If $\rho$ is the \num portfolio, then:

\smallskip

\noindent $\bullet$ it maximizes the \emph{growth rate} over all
portfolios (subsection \ref{subsec: growth-optimal portfolios});

\noindent $\bullet$ it maximizes the \emph{asymptotic growth} of
the wealth process it generates, over all portfolios (Proposition
\ref{prop: asymptotic growth optimality of num});

\noindent $\bullet$ it solves the \emph{relative log-utility
maximization} problem (subsection \ref{subsec: relative
log-optimal}); and

\noindent $\bullet$ $(W^\rho)^{-1}$ minimizes the \emph{reverse
relative entropy} among all \emph{supermartingale deflators}
(subsection \ref{subsec: supermart deflators}).

\end{rem}

We now state the basic problem that will occupy us in this section;
its solution is the content of Theorem \ref{thm: existence of the
num for semimarts}.

\begin{problem} \label{problem: nec and suf pred char}
Find necessary and sufficient conditions for the existence of the
\num portfolio in terms of the triplet of predictable
characteristics of the returns process $X$ (equivalently, of the
stock-price process $S$).
\end{problem}

\subsection{Preliminary necessary and sufficient conditions for existence of the num\'eraire portfolio}
\label{subsec: nec+suf for a port to be the num}

To decide whether $\rho \in \Pi_\K$ is the \num portfolio, we must
check whether $W^\pi / W^\rho$ is a supermartingale for all $\pi \in
\Pi_\K$, so let us   derive a convenient expression for this ratio.

Consider a baseline portfolio $\rho \in \Pi_\K$ that generates a
wealth $W^\rho$, and any other portfolio $\pi \in \Pi_\K$; their
relative wealth process is given by the ratio $\,
 W^\pi / W^\rho =  \Exp (\pi \cdot X) / \Exp (\rho\, \cdot
X)$ from \eqref{eq: wealth process}, which can be further expressed
as follows.

\begin{lem} \label{lem: inverse of stochastic expo}
Suppose that $Y$ and $R$ are two scalar semimartingales with $\Delta
Y > -1$ and $\Delta R > -1$. Then $\Exp(Y) / \Exp(R) = \Exp(Z)$,
where
\begin{equation} \label{eq: inverse of stoch expo}
Z  = Y - R - [Y^\co - R^\co,R^\co] - \sum_{s \leq \cdot}
\set{\Delta (Y_s - R_s) \frac{\Delta R_s}{1 + \Delta R_s}}.
\end{equation}
\end{lem}

\proof The process $\Exp(R)^{-1}$ is locally bounded away from zero,
so the stochastic logarithm $Z$ of $\Exp(Y) / \Exp(R)$ exists.
Furthermore, the process   on the right-hand-side of (\ref{eq:
inverse of stoch expo}) is well-defined and a semimartingale, since
$\sum_{s \leq \cdot} | \Delta R_s |^2 < \infty$ and $\sum_{s \leq
\cdot} |\Delta Y_s \Delta R_s | < \infty$. Now, $\Exp (Y) = \Exp (R)
\Exp(Z) = \Exp(R + Z + [R , Z])$, by Yor's formula. Taking
stochastic logarithms on both sides of the last equation we get $Y =
R + Z + [R,Z]$. This now is an equation for the process $Z$; by
splitting it into continuous and purely discontinuous parts, one can
guess, then verify, that it is solved by the right-hand side of
(\ref{eq: inverse of stoch expo}). \qed

\smallskip

\noindent Using Lemma \ref{lem: inverse of stochastic expo} and
\eqref{eq: wealth process} we get
\[
\frac{W^\pi}{W^\rho} = \Exp
\pare{(\pi - \rho) \cdot X^{(\rho)}}, \quad \hbox{with} \quad
 X^{(\rho)} := X- (c \rho) \cdot G - \bra{\frac{\rho^\top x} {1 +
\rho^\top x} \ x} * \mu\,;
\]
here $\mu$ is the jump measure of $X$   in (\ref{eq: jump measure}),
and $G$ is the operational clock of (\ref{eq: triplet of pred
characteristics}).

\smallskip
We are interested in ensuring that $W^\pi / W^\rho$ is a
supermartingale. Since $W^\pi / W^\rho$ is strictly positive, the supermartingale property is equivalent to the
$\sigma$-supermartingale one, which is in turn equivalent to
requiring that its drift rate be finite and negative. (For drift
rates, see Definition \ref{dfn: drift rate}. For the
$\sigma$-localization technique, see Kallsen \cite{Kallsen:
sigma-localization}; an overview of what is needed here is in
Appendix \ref{subsec: sigma localization}, in particular,
Propositions \ref{prop: pred characterization of (super)marts} and
\ref{prop: pos sigma supermat is supermart}.) Since $W^\pi / W^\rho
= \Exp\pare{(\pi - \rho) \cdot X^{(\rho)}}$, the condition of
negativity on the drift rate of $W^\pi / W^\rho$ is equivalent to
the requirement that the drift rate of the process $(\pi - \rho)
\cdot X^{(\rho)}$ be negative. Straightforward computations show
that, when it exists, this drift rate is
\begin{equation}
\label{eq: rel_perf}
\rel(\pi \such \rho) := (\pi - \rho)^\top b - (\pi - \rho)^\top c
\rho
 + \int \vartheta_{\pi | \rho} (x)\, \nu (\ud x)\,.
\end{equation}
(The notation $\rel(\pi \such \rho)$ stresses that this quantity is
the rate of return of the \emph{relative} wealth process $W^\pi /
W^\rho$.) The integrand $\vartheta_{\pi | \rho} (\cdot)$ in
\eqref{eq: rel_perf} is
\[
\vartheta_{\pi | \rho} (x)\,:=\, \bra{ \frac{(\pi - \rho)^\top
x}{1 + \rho^\top x} - (\pi - \rho)^\top \hx }\,=\, \frac{1 +
\pi^\top x }{1 + \rho^\top x} - 1 - (\pi - \rho)^\top \hx\,;
\]
this satisfies $\nu [x \in \Real^d \such \vartheta_{\pi | \rho} (x)
\leq -1 \textrm{ and } |x| > 1] = 0$, while on $\{ |x| \leq 1 \}$
(near $x=0$) it behaves like $(\rho - \pi)^\top x x^\top \rho$,
comparable to $|x|^2$. The integral in (\ref{eq: rel_perf})
therefore always makes sense, but can take the value $+ \infty$; the
drift rate of $W^\pi / W^\rho$ takes values in $\, \mathbb{R} \cup
\{ + \infty \}\,$,  and the quantity of (\ref{eq: rel_perf}) is
well-defined.

Thus, $W^\pi / W^\rho$ is a supermartingale if and only if $\rel(\pi
\such \rho) \leq 0$, $\prob \otimes G$-almost everywhere. Using this
last fact we get preliminary necessary and sufficient conditions
needed to solve Problem \ref{problem: nec and suf pred char}. In a
different, more general form (involving also ``consumption'') these
have already appeared in Goll and Kallsen \cite{Goll - Kallsen:
log-optimal}.

\begin{lem} \label{lem: necess and suff for numeraire}
Suppose that the constraints $\K$ imply the natural constraints of
\eqref{eq: natural constraints} (i.e., $\K \subseteq \K_0$), and
consider a process $\rho$ with $\rho(\omega, t) \in \K(\omega, t)$
for all $(\omega, t) \in \dbra{0,T}$. This $\rho$ is the \num
portfolio in the class $\Pi_\K$ if and only if:
\begin{enumerate}
    \item $\rel(\pi \such \rho) \leq 0$, $\prob \otimes G$-a.e. for every
    $\pi \in \Pre(\Real^d)$ with $\pi(\omega, t) \in \K(\omega,
    t)$;
    \item $\rho$ is predictable; and
    \item $\rho$ is $X$-integrable.
\end{enumerate}
\end{lem}

\proof The three conditions are clearly sufficient for ensuring that
$W^\pi / W^\rho$ is a supermartingale for all $\pi \in \Pi_\K$.

The necessity is trivial, but for the fact that condition (1) is to
hold not only for all $\pi \in \Pi_\K$, but for \emph{any}
predictable process $\pi$ (not necessarily $X$-integrable) such that
$\pi(\omega, t) \in \K(\omega, t)$. Suppose condition (1) holds for
all $\pi \in \Pi_\K$; first, take any $\xi \in \Pre$ such that $\xi
(\omega, t) \in \K(\omega, t)$ and $\xi^\top \Delta X > -1$. Then
$\xi_n := \xi \indic_{\{ |\xi| \leq n\}} + \rho \indic_{\{ |\xi|
> n\}}$ belongs to $\Pi_\K$, so $\rel(\xi \such \rho)
\indic_{\{ |\xi| \leq n\}} = \rel(\xi_n \such \rho) \indic_{\{ |\xi|
\leq n\}} \leq 0$; sending $n$ to infinity we get $\rel(\xi \such
\rho) \leq 0$. Now pick any $\xi \in \Pre (\Real^d)$ such that $\xi
(\omega, t) \in \K(\omega, t)$; we have $\xi^\top \Delta X \geq -1$
but not necessarily $\xi^\top \Delta X > -1$. Then, for $n \in
\Natural$, $\xi_n := (1-n^{-1}) \xi$ also satisfies $\xi_n \in \Pre
(\Real^d)$ and $\xi_n (\omega, t) \in \K(\omega, t)$ and further
$\xi_n^\top \Delta X > -1$; it follows that $\rel(\xi_n \such \rho)
\leq 0$. Fatou's lemma now gives $\rel(\xi \such \rho) \leq 0$. \qed

\medskip In order to solve Problem \ref{problem: nec and suf pred char},
the conditions of Lemma \ref{lem: necess and suff for numeraire}
will be tackled one by one. For condition (1), it will turn out that
one has to solve for each fixed $(\omega,t) \in \dbra{0,T}$ a convex
optimization problem over the set $\K (\omega, t)$. It is obvious
that if (1) above is to hold for $\K$, then it must also hold for
the closed convex hull of $\K$, so we might as well assume that $\K$
is closed and convex. For condition (2), in order to prove that the
solution we get is predictable, the set-valued process $\K$ must
have some predictable structure; we describe in the next subsection
how this is done. After that, a simple test will give us condition
(3), and we shall be able to provide the solution of Problem
\ref{problem: nec and suf pred char} in Theorem \ref{thm: existence
of the num for semimarts}.

\subsection{The predictable, closed convex structure of
constraints} \label{subsec: structure of pred convex constraints}

We start with a remark concerning \emph{market degeneracies}, i.e.,
linear dependence that some stocks might exhibit at some points of
the base space, causing seemingly different portfolios to produce
the exact same wealth processes; such portfolios should then be
treated as equivalent. To formulate this notion, consider two
portfolios $\pi_1$ and $\pi_2$ with $W^{\pi_1} = W^{\pi_2}$. Take
stochastic logarithms on both sides of the last equality to get
$\pi_1 \cdot X = \pi_2 \cdot X$. Then, $\zeta := \pi_2 - \pi_1$
satisfies $\zeta \cdot X \equiv 0$; this is equivalent to $\zeta
\cdot X^\co = 0, \ \zeta^\top \Delta X = 0$ and $\zeta \cdot B = 0$,
and suggests the following definition.

\begin{defn} \label{dfn: null investments}
For a triplet of predictable characteristics $(b,c,\nu)$, the
linear-subspace-valued process of \textsl{null investments} $\N$ is
the set of vectors (depending on $(\omega, t)$, of course) for which
nothing happens if one invests in them, namely
\begin{equation} \label{eq: null subspace N}
\N(\omega, t) := \big\{\zeta \in \Real^d \such \zeta^\top c(\omega,
t) = 0,\ \nu(\omega, t)[ \zeta^\top x \neq 0 ] = 0  \text{ and }
\zeta^\top b(\omega, t) = 0 \big\}\,.
\end{equation}
\end{defn}

We have $W^{\pi_1} = W^{\pi_2}$ if and only if $\pi_2(\omega, t) -
\pi_1(\omega, t) \in \N (\omega, t)$ for $\prob \otimes G$-almost
every $(\omega, t) \in \dbra{0,T}$; then, the portfolios $\pi_1$ and
$\pi_2$ are considered identical.

\begin{defn} \label{dfn: predictable convex constraints}
The $\Real^d$-set-valued process $\K$ will be said to impose
\textsl{predictable closed convex constraints}, if
\begin{enumerate}
    \item $\N(\omega, t) \subseteq \K(\omega, t)$ for \emph{all}
    $(\omega, t) \in \dbra{0,T}$,
    \item $\K(\omega, t)$ is a closed convex set, for \emph{all}
    $(\omega, t) \in \dbra{0,T}$, and
    \item $\K$ is predictably measurable, in the sense that for any closed $F \subseteq \Real^d$
    we have $ \{ \K \cap F \neq \emptyset \} := \{ (\omega,t) \in \dbra{0,T} \such \K(\omega, t) \cap F \neq \emptyset \} \in
    \Pre\,.$
\end{enumerate}
\end{defn}

Note the insistence that (1), (2) must hold for \emph{every}
$(\omega, t) \in \dbra{0,T}$, not just in an ``almost every'' sense.
Requirement (1) says that we are giving investors \emph{at least}
the freedom to do nothing: if an investment is to lead to absolutely
no profit or loss, one should be free to do it. In the
non-degenerate case this just becomes $0 \in \K(\omega, t)$ for all
$(\omega, t) \in \dbra{0,T}\,$. Appendix \ref{subsec: measurable
selection} discusses further the measurability requirement (3) and
its equivalence with other definitions of measurability.

\smallskip
The \textsl{natural constraints} $\K_0$ of (\ref{eq: natural
constraints}) satisfy the requirements of Definition \ref{dfn:
predictable convex constraints}. For the predictability
requirement, write $\K_0 = \{ \p \in \Real^d \ | \ \int \kappa(1 +
\p^\top x) \nu (\ud x) = 0 \}$, where $\kappa(x) := (x \wedge 0)^2
/ (1+(x \wedge 0)^2)$; then use Lemma \ref{lem: Caratheodory makes
measurable sets} in conjunction with Remark \ref{rem: on
characteristics being nice everywhere}, which provides a version
of the characteristics, such that the integrals in the above
representation of $\K_0$ make sense for \emph{all} $(\omega, t)
\in \dbra{0,T}$. In view of this we shall \emph{always} assume $\K
\subseteq \K_0$, since otherwise we can replace $\,\K$ by $\,\K
\cap \K_0$ (and use the fact that intersections of closed
predictable set-valued processes are also predictable
--- see Lemma \ref{lem: union and intersects of measur set-valued processes} of Appendix
\ref{subsec: measurable selection}).

\subsection{Unbounded Increasing Profit}
\label{subsec: unbounded increasing profit} We proceed with an
effort to understand condition (1) in Lemma \ref{lem: necess and
suff for numeraire}. In this subsection we state a sufficient
condition for its failure in terms of predictable characteristics.
In the next subsection, when we state our first main theorem about
the existence of the \num portfolio, we shall see that this
condition is also necessary. Its failure is intimately related to
the existence of wealth processes that start with unit capital, make
some wealth with positive probability, and are increasing. The
existence of such a possibility in a financial market amounts to the
most egregious form of arbitrage.

\begin{defn} \label{dfn: NUIP}
The predictable set-valued process $\check{\K} := \bigcap_{a > 0} a
\K$ is the set of \textsl{cone points} (or \textsl{recession cone})
of $\K$. A portfolio $\pi \in \Pi_{\Kc}$ will be said to generate an
\textsl{Unbounded Increasing Profit} (UIP), if the wealth process
$W^\pi$ is increasing ($\prob[W^\pi_s \leq W^\pi_t, \forall \ s < t
\leq T] = 1$), and if $\prob[W^\pi_T > 1]
> 0$. If no such portfolio exists, then we say that the \textsl{No
Unbounded Increasing Profit} (NUIP) condition holds.
\end{defn}

The qualifier ``unbounded'' stems from the fact that since $\pi \in
\Pi_{\Kc}$, an agent has unconstrained leverage on the position
$\pi$ and can invest unconditionally; by doing so, the agent's
wealth will be multiplied accordingly. It should be clear that the
\num portfolio cannot exist, if such strategies exist. To obtain the
connection with predictable characteristics, we also give the
definition of the immediate arbitrage opportunity vectors in terms
of the L\'evy triplet.

\begin{defn} \label{dfn: instant_arb_opport}
For a triplet of predictable characteristics $(b,c,\nu)$, the
set-valued process $\I$ of \textsl{immediate arbitrage
opportunities} is defined for any $(\omega, t) \in \Omega \times
\Real_+$ as the set of vectors $\xi \in \Real^d \setminus \N(\omega,
t)$ for which:
\[
\textrm{(1) } \xi^\top c = 0, \ \textrm{ (2) } \nu [ \xi^\top x < 0
] = 0, \ \textrm{ and (3) } \xi^\top b - \int \xi^\top \hx \nu (\ud
x) \geq 0.
\]
(We have hidden the dependance of $(b,c, \nu)$ on $(\omega, t)$
above, to ease the reading.)
\end{defn}

$\N$-valued processes satisfy these three conditions, but cannot be
considered ``arbitrage opportunities'' since they have zero returns.
One can see that $\I$ is cone-valued with the whole ``face'' $\N$
removed.

Assume, for simplicity only, that $X$ is a L\'evy process; and that
we can find a vector $\xi \in \I$ (which is deterministic). In
Definition \ref{dfn: instant_arb_opport}, condition (1) implies that
$\xi \cdot X$ has no diffusion part; (2) implies that $\xi \cdot X$
has no negative jumps; whereas, (3) turns out to imply that $\xi
\cdot X$ has nonnegative drift and is of finite variation (though
this is not as obvious). Using $\xi \notin \N$, we get that $\xi
\cdot X$ is actually non-zero and increasing, and the same will hold
for $W^\xi = \Exp(\xi \cdot X)$; see subsection \ref{subsec: (NIP)
implies I cap K is empty} for a thorough discussion of the general
(not necessarily L\'evy-process) case.

\begin{prop} \label{prop: pred charact of (NUIP)}

The \NUIP condition is equivalent to requiring that the predictable
set $\{\I \cap \check{\K} \neq \emptyset\} := \{(\omega, t) \in
\dbra{0,T} \such \I (\omega, t)  \cap \check{\K} (\omega, t) \neq
\emptyset\}$ be $\prob \otimes G$-null.
\end{prop}

The proof of this result is given in section \ref{sec: The NUIP
condition}. Subsection \ref{subsec: (NIP) implies I cap K is empty}
contains one side of the argument (if an UIP exists, then $\{ \I
\cap \check{\K} \neq \emptyset \}$ \emph{cannot} be $\prob \otimes
G$-null) and makes a rather easy read. The other direction, though
it follows from the same idea, has a ``measurable selection'' flavor
and the reader might wish to skim it.

\begin{rem} \label{rem: if I cap K neq 0, no solution for rel}
We describe briefly the connection between Proposition \ref{prop:
pred charact of (NUIP)} and our original Problem \ref{problem: nec
and suf pred char}. We discuss how \emph{if $\, \I \cap \check{\K}
\neq \emptyset\,$ has non-zero $\prob \otimes G$-measure, one cannot
find a process $\,\rho \in \Pi_\K$ such that $\, \rel( \pi \such
\rho) \leq 0\,$ holds for all} $\,\pi \in \Pi_\K$. Indeed, a
standard measurable selection argument (for details, the reader
should check section \ref{sec: The NUIP condition}) allows us to
infer the existence of a process $\xi$ such that $\xi (\omega, t)
\in \I (\omega, t) \cap \check{\K} (\omega, t)$ on $\{\I \cap
\check{\K} \neq \emptyset \}$ and $\xi = 0$ otherwise. Now, suppose
that $\rho$ satisfies $\rel( \pi \such \rho) \leq 0$, for all $\pi
\in \Pi_\K$. Since $\xi \in \Pi_{\check{\K}}$, we have $n \xi \in
\Pi_\K$ for all $n \in \Natural$, as well as $(1-n^{-1}) \rho + \xi
\in \Pi_\K$ from convexity; but $\K$ is closed-set-valued, so $\rho
+ \xi \in \Pi_\K$. Now from \eqref{eq: rel_perf} and the definition
of $ \mathfrak{I}$, we see that
\[
\rel(\rho + \xi \such \rho) = \ldots = \xi^\top b - \int \xi^\top
\hx \nu(\ud x) + \int  \frac{\xi^\top x}{1 + \rho^\top x} \nu (\ud
x) \ > \ 0
\]
holds on $\{\I \cap \check{\K} \neq \emptyset \}$, which has
positive $\prob \otimes G$-measure. This is a contradiction: there
cannot exist any $\rho \in \Pi_\K$ satisfying $\rel( \pi \such \rho)
\leq 0$ for all $\pi \in \Pi_\K$.

Proving the converse --- namely, if $\{ \I \cap \check{\K} =
\emptyset \}$ is $\prob \otimes G$-null, then one can find a $\rho
\in \Pi_\K$ that satisfies $\rel( \pi \such \rho) \leq 0$ for all
$\pi \in \Pi_\K$ --- is more involved and will be taken on in
section \ref{sec: The Numeraire for General Semimartingales}.
\end{rem}

\begin{ex} \label{ex: (NUIP) for continuous}
If $\nu \equiv 0$, an immediate arbitrage opportunity is a $\xi \in
\Pi_{\Real^d}$ with $c \xi = 0$ and $\xi^\top b > 0$ on a set of
positive $\prob \otimes G$-measure. It follows that if $X$ has
continuous paths, immediate arbitrage opportunities do not exist if
and only if $b$ lies in the range of $c$, i.e., if there exists a
$d$-dimensional process $\rho$ with $b = c \rho$; of course, if $c$
is non-singular $\prob \otimes G$-almost everywhere, this always
holds and $\rho = c^{-1} b$. It is easy to see that this {\it ``no
immediate arbitrage opportunity''} condition is equivalent to $\ud
B_t \ll \ud [X, X]_t$. We refer the reader to Karatzas, Lehoczky and
Shreve \cite{KLS}, Appendix B of Karatzas and Shreve
\cite{Karatzas-Shreve: Math Finance}, and Delbaen and Schachermayer
\cite{DS: FTAP absolutely continuous} for a more thorough
discussion.
\end{ex}

\begin{rem}
Let us write $X = A + M$ for the unique decomposition of a special
semimartingale $X$ into a predictable finite variation part $A$ and
a local martingale $M$, which we further assume is locally
square-integrable. If $\inner{M}{M}$ is the predictable compensator
of $[M,M]$, Example \ref{ex: (NUIP) for continuous} shows that in
continuous-path models the condition for absence of immediate
arbitrage is $\ud A_t \ll \ud \inner{M}{M}_t$. Compare this with the
more complicated way we have defined this notion in Definition
\ref{dfn: instant_arb_opport}. \emph{Could the simple criterion $\ud
A_t \ll \ud \inner{M}{M}_t$ work in more general situations?} It is
easy to see that $\ud A_t \ll \ud \inner{M}{M}_t$ is necessary for
absence of immediate arbitrage opportunities; but it is not
sufficient
--- it is too weak. Take for example $X$ to be the standard Poisson
process. In the non-constrained case, any positive portfolio is an
immediate arbitrage opportunity. Nevertheless, $A_t = t$ and $M_t =
X_t - t$ with $\inner{M}{M}_t = t = A_t$, so that $\ud A_t \ll \ud
\inner{M}{M}_t$ holds trivially.
\end{rem}

\subsection{The growth-optimal portfolio and connection with the num\'eraire portfolio} \label{subsec: growth-optimal portfolios}

We hinted in Remark \ref{rem: if I cap K neq 0, no solution for rel}
that if $\{\I \cap \check{\K} \neq \emptyset\}$ is $\prob \otimes
G$-null, then one can find a process $\rho \in \Pi_\K$ such that
$\rel(\pi \ | \ \rho) \leq 0$ for all $\pi \in \Pi_\K$. It is
actually also important to have a way of computing this process
$\rho$.

\smallskip
For a portfolio $\pi \in \Pi_\K$, its \textsl{growth rate} is
defined as the drift rate of the log-wealth process $\log W^\pi$.
One can use the stochastic exponential formula (\ref{eq: stoch
expo}) and formally (since this will not always exist) compute the
growth rate of $W^\pi$ as
\begin{equation} \label{eq: growth rate}
\g(\pi) := \pi^\top b - \frac{1}{2} \pi^\top c \pi + \int
\bra{\log(1 + \pi^\top x) - \pi^\top \hx} \nu (\ud x).
\end{equation}

We describe (somewhat informally) the connection between the \num
portfolio and the portfolio that maximizes in an $(\omega,
t)$-pointwise sense the growth rate over all portfolios in $\Pi_\K$
in the case of a deterministic triplet. (Note that for the general
semimartingale case this connection has been observed in \cite{Goll
- Kallsen: log-optimal}.) A $\rho \in \K$ maximizes this concave
function $\g$ if and only if the derivative of $\g$ at the point
$\rho$ is negative in all direction $\pi - \rho$, $\pi \in \K$. This
directional derivative is
\[
(\nabla \g)_\rho (\pi - \rho)\,=\, (\pi - \rho)^\top b - (\pi -
\rho)^\top c \rho + \int \bra{ \frac{(\pi - \rho)^\top x}{1 +
\rho^\top x} - (\pi - \rho)^\top \hx } \nu (\ud x),
\]
which is exactly $\rel(\pi \such \rho)$. Of course, we do not know
if we can differentiate under the integral appearing in equation
\ref{eq: growth rate}. Even worse, we do not know a priori whether
the integral is well-defined. Both its positive and negative parts
could lead to infinite results. We now describe a class of L\'evy
measures for which the concave growth rate function $\,\g(\cdot)\,$
of (\ref{eq: growth rate}) \emph{is} well-defined.

\begin{defn} \label{dfn: finite-log-value measure and approximating measures}
A L\'evy measure $\nu$ for which $\int \log(1 + |x|)\, \indic_{\{
|x|
> 1\}} \nu (\ud x) < \infty$ will be said to \textsl{integrate the
log}. Now, consider any L\'evy measure $\nu$; an
\textsl{approximating sequence} is a sequence $(\nu_n)_{n \in
\Natural}$ of L\'evy measures that integrate the log with $\nu_n
\sim \nu$, whose densities $f_n := \ud \nu_n / \ud \nu$ satisfy $0 <
f_n \leq 1$, $f_n (x) = 1$ for $|x| \leq 1$, and $\lim_{n \to
\infty} \uparrow f_n = \indic$.
\end{defn}

There are many ways to choose the sequence $(\nu_n)_{n \in
\Natural}$, or equivalently the densities $(f_n)_{n \in
\Natural}\,$; as a concrete example, take $f_n(x) = \indic_{\{ |x|
\leq 1\}} + |x|^{-1/n} \indic_{\{|x| > 1\}}$.

The integral in (\ref{eq: growth rate}) is well defined and finite,
when the L\'evy measure $\nu$ integrates the log; and then $\rho$ is
the \num portfolio if and only if it maximizes $\g(\cdot)$
pointwise. If $\nu$ fails to integrate the log, we shall consider a
sequence of auxiliary problems as in Definition \ref{dfn:
finite-log-value measure and approximating measures}, then show that
their solutions converge to the solution of the original problem.

\subsection{The first main result}

We are now ready to state the main result of this section, which
solves Problem \ref{problem: nec and suf pred char}. We already
discussed condition (1) of Lemma \ref{lem: necess and suff for
numeraire} and its predictable characterization: \emph{there exists
a predictable process $\rho$ with $\rho (\omega, t) \in \K (\omega,
t)$ such that $\rel(\pi \such \rho) \leq 0$ for all $\pi \in
\Pi_\K$, if and only if $\{\I \cap \check{\K} \neq \emptyset \}$ has
zero $\prob \otimes G$-measure} (Remark \ref{rem: if I cap K neq 0,
no solution for rel}). If this holds, we construct such a process
$\rho$; the only thing that might keep $\rho$ from being the \num
portfolio is failure of $X$-integrability. To deal with this issue,
define for a given predictable $\rho$
\[
\psi^\rho := \nu[\rho^\top x > 1] + \Big | \rho^\top b + \int
\rho^\top x ( \indic_{\{ |x| > 1\}} - \indic_{\{ |\rho^\top x|
> 1 \}} ) \nu (\ud x) \Big |.
\]
Here is the statement of the main result; its proof is given in
section \ref{sec: The Numeraire for General Semimartingales}.

\begin{thm} \label{thm: existence of the num for semimarts}
Consider a financial model described by a semimartingale returns
process $X$ and predictable closed convex constraints $\K$.

\smallskip

\noindent $\bullet$ (1.i) If the predictable set $\{ \I \cap
\check{\K} \neq
            \emptyset \}$ has zero $\prob \otimes G$-measure, then there exists
            a unique process $\rho \in \Pre(\Real^d)$ with $\rho (\omega, t) \in \K \cap
            \N^\bot (\omega, t)$ for all $(\omega, t) \in \dbra{0,T}$, such that
            $\rel(\pi \such \rho) \leq 0$ for all $\pi \in \Pi_\K$.
\smallskip

\noindent $\bullet$ (1.ii) On the predictable set $\Lambda :=
\set{\int\log(1 + |x|) \indic_{\{ |x| > 1 \}}  \nu (\ud x) <
\infty}$, this process $\rho$ is obtained as the unique solution of
the concave optimization problem
            \[
            \rho = \arg \max_{\pi \in \K \cap \N^\bot} \g (\pi)\,.
            \]
In general, $\rho$  can be obtained as the limit of   solutions to
corresponding problems (where one replaces $\nu$ by $ (\nu_n)
 $, an approximating sequence  in the definition of
$\,\g$).

\smallskip

\noindent $\bullet$ (1.iii) Further, if the process $\rho \in
\Pre(\Real^d)$ constructed above is such that $\pare{\psi^\rho \cdot
G}_t < + \infty$ on $\dbra{0, T}\,$, then $\rho$ is $X$-integrable
and   is the \num portfolio.

\medskip

\noindent $\bullet$ 2. Conversely to (1.i)-(1.ii)-(1.iii) above, if
the \num portfolio $\rho$ exists in $\Pi_\K$, then the predictable
set $\{ \I \cap \check{\K} \neq \emptyset \}$ has zero $\prob
\otimes G$-measure, and $\rho$ satisfies $\pare{\psi^\rho \cdot G}_t
< + \infty$ on $\dbra{0, T}\,$, as well as $\, \rel(\pi \such \rho)
\leq 0$ for all $\pi \in \Pi_\K$.

\end{thm}

Let us pause   to comment on the predictable characterization of
$X$-integrability of $\rho$, which amounts to $G$-integrability of
both processes
\begin{equation}
                          \label{eq: phi1 and phi2}
\psi_1^\rho := \nu [ \rho^\top x > 1]~~~~~~ \text{ and } ~~~\
\psi_2^\rho :=  \rho^\top b + \int \rho^\top x \, \big(
\indic_{\{|x|
> 1\}} - \indic_{\{ |\rho^\top x| > 1 \}} \big) \nu (\ud x).
\end{equation}
The integrability of $\psi_1^\rho$ states that $\rho \cdot X$ cannot
have an infinite number of large positive jumps on finite
time-intervals; but this must hold, if $\rho \cdot X$ is to be
well-defined. The second term $\psi_2^\rho$  is exactly the drift
rate of the part of $\rho \cdot X$ that remains when we subtract all
large positive jumps (more than unit in magnitude). This part has to
be a special semimartingale, so its drift rate must be
$G$-integrable, which is exactly the requirement $(\abs{\psi_2^\rho}
\cdot G) < \infty$, on $\dbra{0, T}$.

The requirement $\prob[ (\psi^\rho \cdot G)  < + \infty, \textrm{ on
} \dbra{0, T}] = 1$ does \emph{not} imply $\pare{\psi^\rho \cdot
G}_T < + \infty$ on $\{ T = \infty \}$. The stronger requirement
$\pare{\psi^\rho \cdot G}_T < \infty$ means that $\rho$ is
$X$-integrable up to time $T$, in the terminology of Appendix
\ref{subsec: semimarts up to infty}. This, in turn, is equivalent to
the fact that the \num portfolio exists \emph{and} that $W^\rho_T <
\infty$ (even on $\{ T = \infty \}$). We shall return to this when
we study arbitrage in the next section.

\begin{rem} \label{rem: deterministic functional deciding on num}
The conclusion of Theorem \ref{thm: existence of the num for
semimarts} can be stated succinctly as follows: the \num portfolio
exists if and only if we have $\Psi (B, C, \eta) < \infty$ on
$\dbra{0, T}\,$, for the \emph{deterministic, increasing} functional
$\,\Psi (B, C, \eta) \,:= \, \big( \infty \,\indic_{\{ \I \cap
\check{\K} \neq \emptyset \}} + \psi^\rho\, \indic_{\{ \I \cap
\check{\K} = \emptyset \}} \big) \cdot G\,$ of the triplet of
predictable characteristics $(B, C, \eta)$.
\end{rem}

\begin{ex} \label{ex: (NUIP) for continuous 2}
Consider the unconstrained case $\K = \Real^d$ for the
continuous-path semimartingale case of Example \ref{ex: (NUIP) for
continuous}. Since $(\nabla \g)_\pi = b - c \pi = c \rho - c \pi$ is
trivially zero for $\pi = \rho$, $\rho$ will be the \num portfolio
as long as $\pare{(\rho^\top c \rho) \cdot G} < \infty$ on $\dbra{0,T}$, or, in the
case where $c^{-1}$ exists, when $\pare{(b^\top c^{-1} b) \cdot G}
< \infty$ on $\dbra{0,T}$.
\end{ex}

\subsection{Relative log-optimality} \label{subsec: relative log-optimal}

In this and the next subsection we give two optimality properties of
the \num portfolio. Here we show that it is exactly the {\it
log-optimal portfolio,} if the latter is defined in a relative
sense.

\begin{defn}
A portfolio $\rho \in \Pi_\K$ will be called \textsl{relatively
log-optimal}, if
\[
\expec \bra{ \limsup_{t \uparrow T} \pare{ \log
\frac{W^\pi_t}{W^\rho_t} }} \leq 0 ~ \text{ holds for every } \pi
\in \Pi_\K.
\]
\end{defn}

Here the $\limsup$ is clearly superfluous on $\{ T < \infty \}$ but
we include it to also cover the infinite time-horizon case. If
$\rho$  is relatively log-optimal, the $\limsup$ is actually a
finite limit; this  is an easy consequence of the following result.

\begin{prop} \label{prop: equivalent of numeraire and rel-log-optimal}
A \num portfolio exists if and only if a relatively log-optimal
portfolio exists, in which case the two are the same.
\end{prop}

\proof Whenever we write $W^{\pi_1}_T / W^{\pi_2}_T$ for $\pi_1$ and
$\pi_2$ in $\Pi_\K$, we tacitly imply that on $\{ T = \infty \}$ the
limit of this ratio exists, and   take $W^{\pi_1}_T / W^{\pi_2}_T$
to be that limit.

\smallskip

\noindent $\bullet$ Suppose $\rho$ is a \num portfolio. For any $\pi
\in \Pi_\K$ we have $\expec[W^{\pi}_T / W^{\rho}_T] \leq 1$, and
Jensen's inequality gives $\expec[\log ( W^{\pi}_T/W^{\rho}_T ) ]
\leq 0$, so $\rho$ is also relatively log-optimal.

\smallskip

\noindent $\bullet$ Let us now assume that a relative log-optimal
portfolio $\hat{\rho}$ exists --- we shall show that the \num
portfolio exists and is equal to $\hat{\rho}$. Without loss of
generality, assume that $\hat{\rho} (\omega, t)$ lies on $\N
(\omega, t)$ for $\prob \otimes G$-almost every $(\omega, t) \in
\dbra{0, T}$ --- otherwise, we project $\hat{\rho} (\omega, t)$ on
$\N (\omega, t)$ and observe that the projected portfolio generates
the same wealth as the original.

First, we observe that $\{\I \cap \check{\K} \neq \emptyset\}$
\emph{must} have zero $\prob \otimes G$-measure. To see why, suppose
the contrary. Then, by Proposition \ref{prop: pred charact of
(NUIP)}, we could select a portfolio $\xi \in \Pi_\K$ that leads to
unbounded increasing profit. According to Remark \ref{rem: if I cap
K neq 0, no solution for rel}, we would have
  $\hat{\rho} + \xi \in \Pi_\K$ and $\rel (\hat{\rho} \such
\hat{\rho} + \xi) \leq 0$, with strict inequality  on a predictable
set of positive $\prob \otimes G$-measure; this would mean that the
process $W^{\hat{\rho}} / W^{\hat{\rho} + \xi}$ is a non-constant
positive supermartingale, and Jensen's inequality again would give
$\expec[\log ( W^{\hat{\rho}}_T/W^{\hat{\rho} + \xi}_T ) ] < 0$,
contradicting the relative log-optimality of $\hat{\rho}$.

Continuing, since $\{ \I \cap \check{\K} \neq \emptyset \}$ has
zero $\prob \otimes G$-measure, we can construct the predictable
process $\rho$ which is the candidate in Theorem \ref{thm:
existence of the num for semimarts} (1) for being the \num
portfolio. We only need to show that the predictable set $\{
\hat{\rho}\neq \rho \}$ has zero $\prob \otimes G$-measure. By way
of contradiction, suppose that $A_n := \{ \hat{\rho}\neq \rho, \
|\rho| \leq n \}$ had non-zero $\prob \otimes G$-measure for some
$n \in \Natural$. We then define $\,\pi := \hat{\rho}\, \indic_{
\dbra{0,T} \setminus A_n} + \rho \,\indic_{A_n} \in \Pi_\K\,$ ---
since $\rel(\hat{\rho} \such \pi) = \rel(\hat{\rho} \such \rho)
\indic_{A_n} \leq 0$ with strict inequality on $A_n$, the same
argument as in the end of the preceding paragraph shows that
$\hat{\rho}$ cannot be the relatively log-optimal portfolio. We
conclude that $\{ \hat{\rho}\neq \rho \} = \bigcup_{n \in
\Natural} A_n$ has zero $\prob \otimes G$-measure, and thus $\rho
= \hat{\rho}$ is the \num portfolio. \qed

\medskip

We remark that if the log-utility optimization problem has a finite
value and the condition NFLVR of Delbaen and Schachemayer \cite{DS:
FTAP locally bdd} holds (see also Definition \ref{dfn: arbitrage
notions} below), the result of the last proposition is well-known
--- see for example Kramkov and Schachermayer \cite{Kramkov-Schachermayer: Asymptotic elast and util
function}. Christensen and Larsen \cite{Chr-Lar: NA and GOP} start
by adopting the above ``relative'' definition as log-optimality (or,
as they call it ``growth optimality'') and eventually show that the
growth-optimal is equal to the \num portfolio.

\begin{ex}
Take a one-stock market model with $S_t = \exp(\beta_{T \wedge
t})$, where $\beta$ is a standard, one-dimensional Brownian motion
and $T$ an a.s. finite stopping time with $\expec \bra{\beta^-_T}
< + \infty$ and $\expec \bra{\beta^+_T}
= + \infty$. Then  $\expec \bra{\log S_T} = + \infty$ and the
classical log-utility optimization problem is not well-posed (one
can find a multitude of portfolios with infinite expected
log-utility). In this case, Example \ref{ex: (NUIP) for
continuous} shows that $\rho = 1/2$ is both the \num and the
relative log-optimal portfolio.
\end{ex}

\subsection{An asymptotic optimality property}

In this subsection we deal with a purely infinite time-horizon case
$T \equiv \infty$ and   describe an ``asymptotic growth optimality''
property of the \num portfolio $\rho$, which we assume it exists.
Note that for any portfolio $\pi \in \Pi_\K$ the process $W^\pi /
W^\rho$ is a positive supermartingale and therefore the $\lim_{t \to
\infty} (W^\pi_t / W^\rho_t)$ exists in $\,[0, +\infty)$.
Consequently, for \emph{any} increasing process $H$ with $H_\infty =
+ \infty$ ($H$ does not even have to be adapted), we have
$\limsup_{t \to \infty} \left( (H_t)^{-1} \log (W^\pi_t / W^\rho_t)
\right)\, \leq \, 0$. A closely-realted version of ``asymptotic
growth optimality'' was first observed and proved in Algoet and
Cover \cite{Algoet-Cover} for the discrete-time case; see also
Karatzas and Shreve \cite{Karatzas-Shreve: Math Finance} and Goll
and Kallsen \cite{Goll - Kallsen: log-optimal} for a discussion of
this asymptotic optimality in the continuous-path and the general
semimartingale case, respectively. In the above-mentioned works, the
authors prove that $\limsup_{t \to \infty} \left( t^{-1} \log
W^\pi_t \right) \leq \limsup_{t \to \infty} \left( t^{-1} \log
W^\rho_t \right)\, \leq \, 0$, which is certainly a weaker statement
than what we mention (interestingly, the proof used is more
involved, using a ``Borel-Cantelli''-type argument).

Our next result, Proposition \ref{prop: asymptotic growth optimality
of num},  separates the cases when $\lim_{t \to \infty} (W^\pi_t /
W^\rho_t)$ is $(0, \infty)$-valued and when it is zero, and
describes this dichotomy in terms of predictable characteristics. In
the case of convergence to zero, it quantifies how fast this
convergence takes place. Its proof is given in section \ref{sec:
Rates of Conv to Zero of Pos Supermarts}.

\begin{prop} \label{prop: asymptotic growth optimality of num}
Assume that the \num portfolio $\rho$ exists on $\dbra{0,\infty}$.
For any other $\pi \in \Pi_\K$, define the positive, predictable
process
\[
h^{\pi} := - \rel(\pi \such \rho) + \frac{1}{2} (\pi - \rho)^\top
c (\pi - \rho) + \int q_a
\pare{\frac{1+ \pi^\top x}{1+ \rho^\top x}} \nu( \ud x).
\]
and the increasing, predictable process $H^\pi := h^\pi \cdot G$.
Here we use the positive, convex function $q_a(y) := \bra{-\log a +
(1-a^{-1}) y} \indic_{[0, a)} (y) + \bra{y - 1 - \log y}
\indic_{[a,+\infty)} (y)$ for some $a \in (0,1)$. Then, on
$\{H^{\pi}_\infty < + \infty\}$, $\lim_{t \to \infty}
  (W^\pi_t / W^\rho_t ) \, \in \, (0, +\infty)$, while
\[
  \text{on } \{H^{\pi}_\infty = + \infty\}, \quad \limsup_{t \to \infty}
 \left(  \frac{1}{H^\pi_t} \log \frac{W^\pi_t}{W^\rho_t} \right) \, \leq \,
  -1.
\]
\end{prop}

\smallskip
\begin{rem}
Some comments are in order. We begin with the ``strange-looking''
function $q_a (\cdot)$, that depends also on the (cut-off point)
parameter $a \in (0,1)$. Ideally we would like to define $q_0 (y) =
y-1-\log y$ for all $y > 0$, since then the predictable increasing
process $H^\pi$ would be exactly the negative of the drift of the
semimartingale $\log(W^\pi / W^\rho)$. Unfortunately, a problem
arises when the positive predictable process $\int q_0
\pare{\frac{1+ \pi^\top x}{1+ \rho^\top x}} \nu( \ud x)$ fails to be $G$-integrable, which
is equivalent to saying that $\log(W^\pi / W^\rho)$ is \emph{not} a
special semimartingale; the problem comes from the fact that
$q_0(y)$ explodes to $+ \infty$ as $y \downarrow 0$. For this
reason, we define $q_a(\cdot)$ to be equal to $q_0(\cdot)$ on $[a,
\infty)$, linear on $[0,a)$, and continuously differentiable at the
``gluing'' point $a$. The functions $q_a (\cdot)$ are all
finite-valued at $y=0$ and satisfy $q_a (\cdot) \uparrow q_0
(\cdot)$ as $a \downarrow 0$.

Let us now study $h^\pi$ and $H^\pi$. Observe that $h^\pi$ is
\textsl{predictably convex} in $\pi$, namely, if $\pi_1$ and
$\pi_2$ are two portfolios and $\lambda$ is a $[0,1]$-valued
predictable process, then $h^{\lambda \pi_1 + (1-\lambda) \pi_2}
\leq \lambda h^{\pi_1} + (1 - \lambda) h^{\pi_2}$. This, together
with the fact that $h^\pi = 0$ if and only if $\pi - \rho$ is a
null investment, casts $h^\pi$ as a \emph{measure of instantaneous
deviation of $\pi$ from} $\rho$; by the same token, $H^\pi_\infty$
can be seen as the total (cumulative) deviation of $\pi$ from
$\rho$. With this in mind, Proposition \ref{prop: asymptotic
growth optimality of num} says that, if an investment deviates a
lot from the \num portfolio $\rho$ (i.e., if $H^\pi_\infty = +
\infty$), its long-term performance will lag considerably behind
that of $\rho$. Only if an investment tracks very closely the \num
portfolio over $[0, \infty)$ (i.e., if $H^\pi_\infty < + \infty$)
will the two wealth processes have comparable growth. Letting $a
\downarrow 0$ in the definition of $H^\pi$ we get equivalent
measures of distance of a portfolio $\pi$ from the \num portfolio
because $\{ H^\pi_\infty = + \infty \}$ does not depend on the
choice of $a$; nevertheless we get ever sharper results, since
$h^\pi$ is increasing for decreasing $a \in (0,1)$.
\end{rem}

 \smallskip

\section{Unbounded Profits with Bounded Risks, Supermartingale Deflators, and the Num\'eraire Portfolio} \label{sec: Arb, ESMM and supermart defl}

In this section we proceed to investigate how the existence or
non-existence of the \num portfolio relates to some concept of
``free lunch" in the financial market. We shall eventually prove a
version of the Fundamental Theorem of Asset Pricing; this is our
second main result, Theorem \ref{thm: num iff def-non-empty iff
NUPBR}.

\subsection{Arbitrage-type definitions}

We first recall two widely known no-free-lunch conditions for
financial markets (NA and the stronger NFLVR), together with yet
another notion which is exactly what one needs to bridge the gap
between the previous two, and will actually be the most important
for our discussion.

\begin{defn} \label{dfn: arbitrage notions}
For the following definitions we consider our financial model with
constrains $\K$ on portfolios. When we write $W^{\pi}_T$ for some
$\pi \in \Pi_\K$ we assume tacitly that $\lim_{t \to \infty}
W^\pi_t$ exists on $\{ T = \infty \}$, and set $W^\pi_T$ equal to
that limit.

\smallskip

\noindent $\bullet$ A portfolio $\pi \in \Pi_\K$ is said to generate
an \textsl{arbitrage opportunity}, if $\prob[W^\pi_T \geq 1] = 1$
and $\prob[W^\pi_T > 1] > 0$. If no such portfolio exists we say
that the $\K$-constrained market satisfies the \textsl{No Arbitrage}
condition, which we denote by NA$_{\K}$.

\smallskip

\noindent $\bullet$ A sequence $(\pi_n)_{n \in \Natural}$ of
portfolios in $\Pi_\K$ is said to generate an \textsl{unbounded
profit with bounded risk} (UPBR), if the collection of positive
random variables $(W^{\pi_n}_T)_{n \in \Natural}$ is unbounded in
probability, i.e., if $\, \downarrow \lim_{m \to \infty} \sup_{n \in
\Natural}\, \prob [W^{\pi_n}_T > m] > 0$. If no such sequence
exists, we say that the constrained market satisfies the \textsl{no
unbounded profit with bounded risk} (NUPBR$_\K$) condition.

\smallskip

\noindent $\bullet$ A sequence $(\pi_n)_{n \in \Natural}$ of
portfolios in $\Pi_\K$ is said to be a \textsl{free lunch with
vanishing risk} (FLVR), if there exist an $\epsilon > 0$ and an
increasing sequence $(\delta_n)_{n \in \Natural}$ with $0 \leq
\delta_n \uparrow 1$, such that $\prob[W^{\pi_n}_T > \delta_n] =
1$ as well as $\prob[W^{\pi_n}_T > 1 + \epsilon] \geq \epsilon$.
If no such sequence exists, we say that the market satisfies the
\textsl{no free lunch with vanishing risk} (NFLVR$_{\K}$)
condition.
\end{defn}

The NFLVR condition was introduced by Delbaen and Schachermayer
\cite{DS: FTAP locally bdd} in a slightly different form. With the
above definition of FLVR and the convexity Lemma A 1.1 from
\cite{DS: FTAP locally bdd}, we can further assume that there exists
a $[1, + \infty]$-valued random variable $f$ with $\prob[f
> 1] > 0$ such that $\prob$-$\lim_{n \to \infty} W^{\pi_n}_T = f$;
this brings us back to the usual definition in \cite{DS: FTAP
locally bdd}.

If an UPBR exists, one can find a sequence of wealth processes, each
starting with less and less capital (converging to zero) and such
that the terminal wealths are unbounded with a fixed probability.
Thus, UPBR can be translated as ``the \emph{possibility} of making
(a considerable) amount out of almost nothing''; it should be
contrasted with the classical notion of arbitrage, which can be
translated as ``the \emph{certainty} of making something more out of
something''.

Observe that \NUPBRC can be alternatively stated by using portfolios
with \emph{bounded support}, so   the requirement of a limit at
infinity for the wealth processes on $\{ T = \infty \}$ is
automatically satisfied. This is relevant   because, as we shall
see, when \NUPBRC holds every wealth process $W^\pi$ has a limit on
$\{ T = \infty \}$ and   is a semimartingale up to $T$ in the
terminology of Appendix \ref{subsec: semimarts up to infty}.

\smallskip
None of the two conditions \NAC and \NUPBRC implies the other, and
they are not mutually exclusive. It is easy to see that they are
both weaker than NFLVR$_\K$, and that in fact we have the following
result which gives the exact relationships among these notions under
cone constraints. Its proof can be found in \cite{DS: FTAP locally
bdd} for the unconstrained case; we include it here for
completeness.

\begin{prop} \label{prop: (NFLVR) equiv to (NA) + bdd in prob}
Suppose that $\K$ enforces predictable closed convex \emph{cone}
constraints. Then, \emph{NFLVR}$_\K$ holds, if and only if both
\emph{NA}$_\K$ and \emph{NUPBR}$_\K$ hold.
\end{prop}

\proof It is obvious that if either \NAC or \NUPBRC fail, then
\NFLVRC fails too. Conversely, suppose that \NFLVRC fails. If \NAC
fails there is nothing more to say, so suppose that \NAC holds and
let $(\pi_n)_{n \in \Natural}$ generate a free lunch with
vanishing risk. Under \NAC, the assumption $\prob[W^{\pi_n}_T >
\delta_n] = 1$ results in the stronger $\prob(W^{\pi_n}_t >
\delta_n \textrm{ for all } t \in [0, T]) = 1$. Construct a new
sequence of wealth processes $(W^{\xi_n})_{n \in \Natural}$ by
requiring $W^{\xi_n} = 1 + (1-\delta_n)^{-1} (W^{\pi_n} -1)$,
check that $W^{\xi_n} > 0$, and then that $\xi_n \in \Pi_\K$ (here
it is essential that $\K$ be a cone). Furthermore,
$\prob[W^{\pi_n}_T \geq 1 + \epsilon] \geq \epsilon$ becomes
$\prob[W^{\xi_n}_T
> 1 + (1-\delta_n)^{-1} \epsilon] \geq \epsilon\,$; thus
$(\xi_n)_{n \in \Natural}$ generates an unbounded profit with
bounded risk and  \NUPBRC fails. \qed

\subsection{  Fundamental Theorem of Asset Pricing (FTAP)}

The \NFLVRC condition has proven very fruitful in contexts where we
can change the original probability measure $\prob$ to some other equivalent
probability measure $\qprob$, under which the wealth processes have some
kind of (super)martingale property.
\begin{defn} \label{dfn: equivalnet supermart measure}
Consider a financial market model described by a semimartingale
discounted stock price process $S$ and predictable closed convex
constraints $\K$ on portfolios. A probability $\qprob$ will be
called a \textsl{equivalent $\K$-supermartingale measure} (\ESMMC
for short), if $\qprob \sim \prob$ on $\F_T$, and if $W^\pi$ is a
$\qprob$-supermartingale for every $\pi \in \Pi_\K$. The class of
\ESMMC is denoted by $\QprobC$.

Similarly, define a \textsl{equivalent $\K$-local martingale
measure} (\ELMMC for short) $\qprob$ by requiring $\qprob \sim
\prob$ on $\F_T$ and that $W^\pi$ be a $\qprob$-local martingale for
every $\pi \in \Pi_\K$.
\end{defn}

In Definition  \ref{dfn: equivalnet supermart measure} we might as
well assume that $\K$ are cone constraints; because, if \ESMMC
holds, the same holds for the market under constraints
$\overline{\cone} (\K)$, the closure of the \emph{cone generated by}
$\K$.

The following result is one of the best-known in mathematical
finance; we present its ``cone-constrained'' version.

\begin{thm} \label{thm: FTAP}
\textsc{(FTAP) } For a financial market model with stock-price
process $S$ and predictable closed convex cone constraints $\K$,
\emph{NFLVR}$_\K$ is equivalent to $\QprobC \neq \emptyset$.
\end{thm}

Because we are working under constraints, we cannot hope in general
for anything better than an equivalent \emph{supermartingale}
measure in the statement of Theorem \ref{thm: FTAP}. One can see
this easily in the case where $X$ is a single-jump process which
jumps at a stopping time $\tau$ with $\Delta X_\tau \in (-1,0)$ and
we are constrained in the cone of positive strategies. Under any
measure $\qprob \sim \prob$, the process $S = \Exp(X) = W^1$, an
admissible wealth process, will be non-increasing and not
identically zero; this prevents it from being even a local
martingale.

The implication $\QprobC \neq \emptyset \ \Rightarrow$ \NFLVRC is
easy; the reverse is considerably harder for the general
semimartingale model. Several papers are devoted to proving some
version of Theorem \ref{thm: FTAP}; in the generality assumed here,
a proof appears in Kabanov \cite{Kabanov: FTAP}, although all the
crucial work was done by Delbaen and Schachermayer \cite{DS: FTAP
locally bdd} and the theorem is certainly due to them. Theorem
\ref{thm: FTAP} can be derived from Kabanov's statement, since the
class of wealth processes $(W^\pi)_{\pi \in \Pi_\K}$ is convex and
closed in the semimartingale (also called ``\'Emery'') topology. A
careful inspection in M\'emin's work \cite{Memin: semimartingales}
of the proof that the set of all stochastic integrals with respect
to the $d$-dimensional semimartingale $X$ is closed under this
topology, shows that one can pick the limiting semimartingale from a
convergent sequence $(W^{\pi_n})_{n \in \Natural}$, with $\pi_n \in
\Pi_\K$ for all $n \in \Natural\,$, to be of the form $W^\pi$ for
some $\pi \in \Pi_\K\,$.

\subsection{Beyond the Fundamental Theorem of Asset Pricing} \label{subsec: beyond FTAP}

Let us study some   more   the assumptions and   statement of
Theorem \ref{thm: FTAP}. We shall be concerned with three
questions, which will turn out to have the same answer; this
answer will be linked with the \NUPBRC condition and --- as we
shall see in Theorem \ref{thm: num iff def-non-empty iff NUPBR}
--- with the existence of the \num portfolio.

\subsubsection{Convex but non-conic constraints} \label{subsubsec: convex non-cone constraints}
In the statement of Theorem \ref{thm: FTAP} it is \emph{crucial}
that the constraint be a \emph{cone} --- the result fails without
the ``cone'' assumption. Of course, $\QprobC \neq \emptyset
\Rightarrow$ \NFLVRC still holds, but the reverse does not, as
shown in the example below (a raw version of a similar example
from \cite{K: arbitrage for Levy}).

\begin{ex}
Consider two stocks with discounted prices $S^1$ and $S^2$ in a
simple one-period, discrete-time model. We have $S^1_0 = S^2_0 =
1$, while $S^1_1 = 1 + e$ and $S^2_1 = f$. Here $e$ and $f$ are
two independent, exponentially distributed random variables. The
class of portfolios is easily identified with all $(p, q)\in \K_0
= \Real_+ \times [0,1]$. Since $X^1_1 = S^1_1 - S^1_0 = e
> 0$, $\prob$-a.s., we have that NA fails for this (unconstrained)
market. In other words, for the non-constrained case there can be no
ESMM.

Consider now the non-random constraint set $\K = \{(p,q) \in \K_0
\such p^2 \leq q \}$. Observe that $\overline{\cone} (\K) = \Real_+
\times \Real$ and thus no \ESMMC exists; for otherwise
 an ESMM would exist already for the unconstrained case. We
shall nevertheless show in the following paragraph that \NFLVRC
holds for this constrained market.

For a sequence of portfolios $\pi_n \equiv (p_n, q_n)_{n \in
\Natural}$ in $\K$, the wealth on day one will be $W^{\pi_n}_1 = 1 -
q_n + q_n f + p_n e$; obviously $\prob[W^{\pi_n}_1 \geq 1 - q_n] =
1$, since $1 - q_n$ is the essential infimum of $W^{\pi_n}_1$. It
then turns out that in order for $(\pi_n)_{n \in \Natural}$ to
generate a FLVR we must require $q_n \downarrow 0$ and
$\prob[W^{\pi_n}_1 > 1 + \epsilon]
> \epsilon$ for some $\epsilon
> 0$. Observe that we \emph{must} have $q_n > 0$, otherwise $p_n = 0$ as well
(because of the constraints) and then $W^{\pi_n}_1 = 1$. Now,
because of the constraints again we have $|p_n| \leq \sqrt{q_n}$;
since $\prob[e > 0] = 1$ the sequence of strategies $\xi_n :=
(\sqrt{q_n}, q_n)$ will generate a sequence of wealth processes
$(W^{\xi_n})_{n \in \Natural}$ that will dominate $(W^{\pi_n})_{n
\in \Natural}$: $\prob[W_1^{\xi_n} \geq W_1^{\pi_n}] = 1$; this will
of course mean that $(W^{\xi_n})_{n \in \Natural}$ is also a FLVR.
We should then have $\prob[1 - q_n + \sqrt{q_n} e + q_n f > 1 +
\epsilon] > \epsilon$; using $q_n > 0$ and some algebra we get
$\prob[e > \sqrt{q_n} (1 - f) + \epsilon / \sqrt{q_n}] > \epsilon$.
Since $(q_n)_{n \in \Natural}$ goes to zero this would imply that
$\prob[e > M] \geq \epsilon$ for all $M > 0$, which is clearly
ridiculous. We conclude that \NFLVRC holds, although as we have seen
$\QprobC = \emptyset$.
\end{ex}

What can we say then in the case of convex --- but non necessarily
conic --- constraints? It will turn out that for the equivalent of
the FTAP, the assumptions from \emph{both} the economic \emph{and}
the mathematical side should be relaxed. The relevant economic
notion will be \NUPBRC and the mathematical one will be the concept
of supermartingale deflators
--- more on this in subsections \ref{subsec: supermart deflators} and
\ref{subsec: second main result}.

\subsubsection{Describing Free Lunches in terms of Predictable
Characteristics} \label{subsubsec: free lunches and pred char}

The reason why ``free lunches''  are considered economically unsound
stems from the following reasoning: if they exist in a market, many
agents will try to take advantage of them; then, usual
supply-and-demand arguments will imply that some correction on the
prices of the assets will occur, and remove these kinds of
opportunities. This is a very reasonable line of thought,
\emph{provided that one can discover the free lunches that are
present}. But is it true that, given a specific model, one is in a
position  to decide whether free lunches exist or not? In other
words, mere knowledge of the \emph{existence} of a free lunch may
not be enough to carry the previous economic argument
--- one should be able to \emph{construct} the free lunch. This goes
somewhat hand in hand with the fact that the FTAP is a \emph{pure
existence} result, in the sense that it provides knowledge that
\emph{some} equivalent (super)martingale measure exists; in some
cases one might be able to spot it, in other cases not.

A natural question arises: \emph{when free lunches exist, is there a
way to construct them from the predictable characteristics of the
model?} Here is an answer: if \NUPBRC fails, then an UPBR can be
constructed using the triplet $(B, C, \eta)$. The detailed statement
will be given in subsection \ref{subsec: what happens when num fails
to exist}, but let us say here that the \emph{deterministic}
positive functional $\Psi$ of Remark \ref{rem: deterministic
functional deciding on num} is such that on the event $\{\Psi_T
(B,C,\eta) = \infty\}$ \NUPBRC fails (and then we can construct free
lunches using the predictable characteristics), while on $\{\Psi_T
(B,C,\eta) < \infty\}$ \NUPBRC holds. As a result, we see that
\NUPBRC is somehow a \emph{pathwise} notion.

What we described in the last paragraph for the \NUPBRC condition
does not apply to the \NAC condition, as we demonstrate in Example
\ref{ex: no pred char of arb}.

\begin{ex} \label{ex: Bessel}
\textsc{Arbitrage for the Three-Dimensional Bessel Process}.
Consider a one-stock market on the finite time horizon $[0,1]$, with
$S_0 = 1$ and $S$ satisfying the stochastic differential equation
$\ud S_t = (1/S_t) \ud t + \ud \beta_t$. Here, $\beta$ is a
standard, one-dimensional Brownian motion, so $S$ is the
\textsl{three-dimensional Bessel process}. Writing $\ud S_t / S_t =
(1/S^2_t) \ud t + (1/S_t) \ud \beta_t =: \ud X_t$ and using Example
\ref{ex: (NUIP) for continuous 2}, the \num portfolio for the
unconstrained case exists and is $\rho = 1$.

This market admits arbitrage. To wit, with the notation
\[
\Phi(x) = \int_{- \infty}^x \frac{e^{-u^2/2}}{\sqrt{2 \pi}} \ud
u\,,~~~ \ F(t, x) = \frac{\Phi(x / \sqrt{1 - t})}{\Phi(1)}\,, \
\text{ for } x \in \Real \textrm{ and } 0 < t < 1,
\]
consider the process $W_t = F(t, S_t)$. Obviously $W_0 = 1$, $W >
0$ and
\[
\ud W_t = \frac{\partial F}{\partial x} (t, S_t) \ud S_t,
\textrm{~~ and thus } \frac{\ud W_t}{W_t} = \bra{\frac{1}{F(t,
S_t)} \frac{\partial F}{\partial x} (t, S_t)} \ud S_t
\]
by It\^o's formula. We   conclude that $W = W^\pi$ for $\pi_t :=
(\partial \log F /
\partial x) (t, S_t)$, and that $W^\pi_1 = 1 / \Phi(1)
> 1$, i.e., there exists arbitrage in the market.

We remark that there is  also an indirect way to show that arbitrage
exists using the FTAP, proposed by Delbaen and Schachermayer
\cite{DS: Bessel}; there, one has to \emph{further} assume that the
filtration $\mathbf{F}$ is the one generated by $S$ (equivalently,
by $\beta$).
\end{ex}

\emph{This is one of the rare occasions, when one can compute the
arbitrage portfolio concretely}. We were successful in this, because
of the very special structure of the three-dimensional Bessel
process; every model has to be attacked in a different way, and
there is no general theory that will spot the arbitrage.
Nevertheless, we refer the reader to Fernholz, Karatzas and Kardaras
\cite{FKK} and Fernholz and Karatzas \cite{FK} for many examples of
arbitrage relatively to the \emph{market portfolio} (whose wealth
process follows exactly the index $\sum_{i=1}^d S^i$ in proportion
to the initial investment). This is done under conditions on market
structure that are easy to check, and descriptive  -- as opposed to
normative, such as ELMM.

\smallskip

We now show that there \emph{cannot} exist a deterministic positive
functional $\Psi$ that takes for its arguments triplets of
predictable characteristics such that NA holds whenever
$\prob[\Psi_T (B,C,\eta) < \infty] = 1$. Actually, we shall
construct in the next paragraph two stock-price processes on the
\emph{same} stochastic basis and with the \emph{same} predictable
characteristics, and such that NA fails with respect to the one but
holds with respect to the other.

\begin{ex} \label{ex: no pred char of arb}

\textsc{No  Predictable Characterization of Arbitrage.} Assume that
$(\Omega, \F, \prob)$ is rich enough to accommodate two independent
standard one-dimensional Brownian motions $\beta$ and $\gamma$; the
filtration will be the (usual augmentation of the) one generated by
the pair $ (\beta , \gamma) $. We work in the time-horizon $[0,1]$.
Let $R$ be the three-dimensional Bessel process with $R_0 = 1$ and
$\ud R_t = (1/R_t) \ud t + \ud \beta_t$. As $R$ is adapted to the
filtration generated by $\beta$, it is independent of $\gamma$.
Start with the market described by the stock-price $S = R$; the
triplet of predictable characteristics $(B, C, \eta)$ consists of
$B_t = C_t = \int_0^t (1/R_u)^2 \ud u$ and $\eta = 0$. According to
Example \ref{ex: Bessel}, NA fails for this market.

With the same process $R\,$, define now a new stock $\widehat{S}$
following the dynamics $\ud \widehat{S}_t / \widehat{S}_t =
(1/R_t)^2 \ud t + (1/R_t) \ud \gamma_t$ with $\widehat{S}_0 = 1$.
The new dynamics involve $\gamma$, so $\widehat{S}$ is \emph{not} a
three-dimensional Bessel process; nevertheless, it has exactly the
same triplet of predictable characteristics as $S$. But now NA holds
for the market that consists of the stock $\widehat{S}$. We can
actually construct an ELMM, since the independence of $R$ and
$\gamma$ imply that the exponential local martingale $Z := \Exp(-(1
/ R) \cdot \gamma)$ is a \emph{true} martingale; Lemma \ref{lem:
girsanov for independent} below will show this. We can then define
$\qprob \sim \prob$ via $\ud \qprob / \ud \prob = Z_1$, and
Girsanov's theorem will imply that $\widehat{S}$ is the stochastic
exponential of a Brownian motion under $\qprob$ --- thus a true
martingale.
\end{ex}

\begin{lem} \label{lem: girsanov for independent}
On a stochastic basis $(\Omega, \F, \filtration, \prob)$ let $\beta$
be a standard one-dimensional $\mathbf{F}$-Brownian motion, and
$\alpha$   a predictable process, \emph{independent} of $\beta$,
that satisfies $\int_0^t |\alpha_u|^2 \ud u < \infty$, $\prob$-a.s.
Then, the exponential local martingale $Z = \Exp(\alpha \cdot
\beta)$ satisfies $\expec [Z_t] = 1$, i.e., is a true martingale on
$[0,t]$.
\end{lem}

\proof We begin by enlarging the filtration to $\mathbf{G}$ with
$\G_t := \F_t \vee \sigma (\alpha_t; t \in \Real_+)$, i.e., we throw
the \emph{whole} history of $\alpha$ up to the end of time in
$\mathbf{F}$. Since $\alpha$ and $\beta$ are independent, it is easy
to see that $\beta$ is a $\mathbf{G}$-Brownian motion. Of course,
$\alpha$ is a $\mathbf{G}$-predictable process and thus the
stochastic integral $\alpha \cdot \beta$ is the same seen under
$\mathbf{F}$ or $\mathbf{G}$. Then, with $A_n := \{ n-1 \leq
\int_0^t |\alpha_u|^2 \ud u < n \} \in \G_0$ and in view of
$\expec[Z_t \,|\, A_n] = 1$ (since on $A_n$ the quadratic variation
of $\alpha \cdot \beta$ is bounded by $n$), we have $\expec[Z_t] =
\expec[\,\expec[Z_t \,|\, \G_0]\,] = \sum_{n = 1}^\infty
\expec[\,Z_t \,|\, A_n\,] \, \prob[A_n] = 1$. \qed

\subsubsection{Connection with utility maximization} \label{subsubsec: connection with util maxim}

A central problem of mathematical finance is the maximization of
\emph{expected utility from terminal wealth} of an economic agent
who can invest in the market. The agent's preferences are described
by a \textsl{utility function}: namely,  a \emph{concave} and
\emph{strictly increasing} function $U : (0, \infty) \mapsto \Real$.
We also define $U(0) \equiv U(0+)$ by continuity. Starting with
initial capital $w > 0$, the objective of the investor is to find a
portfolio $\hat{\rho} \equiv \hat{\rho} (w) \in \Pi_\K$ such that
\begin{equation} \label{eq: utility optimization}
\expec [ U(w W^{\hat{\rho}}_T) ]  = \sup_{\pi \in \Pi_\K} \expec
\bra{U(w W^{\pi}_T)} =: u(w).
\end{equation}

Probably the most important example is the logarithmic utility
function $U(w) = \log w$. Due to this special structure, when the
optimal portfolio exists it does not depend on the initial capital,
or on the given time-horizon $T$ (``myopia"). We saw in subsection
\ref{subsec: relative log-optimal} that under a suitable
reformulation of log-optimality, the two notions of log-optimal and
\num portfolio are equivalent.

We consider here utility maximization from terminal wealth that is
constrained to be positive (in other words, $U(w) = - \infty$ for $w
< 0$). This problem has a long history; it has been solved in a very
satisfactory manner for general semimartingale models using
previously-developed ideas of martingale duality by Kramkov and
Schachermayer \cite{Kramkov-Schachermayer: Asymptotic elast and util
function, Kramkov-Schachermayer: nec and suf incomplete markets},
where we send the reader for further details.

A common assumption in this context is that the class of equivalent
local martingale measures is non-empty, i.e., that NFLVR holds.
(Interestingly, in Karatzas, Lehoczky, Shreve and Xu \cite{KLSX}
this assumption is \emph{not} made.) The three-dimensional Bessel
process Example \ref{ex: Bessel} shows that this is not necessary;
indeed, since the \num portfolio $\rho = 1$ exists and $\expec [\log
S_1 ] < \infty$, Proposition \ref{prop: equivalent of numeraire and
rel-log-optimal} shows that $\rho$ is the solution to the
log-utility optimization problem. Nevertheless, we have seen that
NFLVR fails for this market. To wit: an investor with log-utility
will optimally choose to hold the stock and, \emph{even though
arbitrage opportunities exist in the market, the investor's optimal
choice is clearly \emph{not} an arbitrage}.

\smallskip

In the mathematical theory of economics, the equivalence of no free
lunches, equivalent martingale measures, and existence of optimal
investments for utility-based preferences, is something of a
``folklore theorem''. Theorem \ref{thm: FTAP} deals with the
equivalence of the first two of these conditions, but the
three-dimensional Bessel process example shows that this does not
completely cover minimal conditions for utility maximization; in
that example, although NA fails, the \num and log-optimal portfolios
do exist. In Theorem \ref{thm: num iff def-non-empty iff NUPBR} we
shall see that \emph{existence of the \num portfolio is equivalent
to the \NUPBR condition}, and in subsection \ref{subsec: Utility
optimization}    that NUPBR is actually the \emph{minimal} ``no free
lunch''-type notion needed to ensure existence of solution to
\emph{any} utility maximization problem. In a loose sense (to become
precise there) the problem of maximizing expected utility from
terminal wealth is solvable for a rather large class of utility
functions, \emph{if and only if} the special case of the logarithmic
utility problem has a solution ---  which is exactly when NUPBR
holds. Accordingly, the existence of an equivalent (local)
martingale measure will have to be substituted by the weaker
requirement, the existence of a \textsl{supermartingale deflator},
which is the subject of the next subsection.

\subsection{Supermartingale deflators} \label{subsec: supermart deflators}

In the spirit of Theorem \ref{thm: FTAP}, we would like now to find
a mathematical condition equivalent to NUPBR. The next concept,
closely related to that of equivalent supermartingale measures but
weaker, will be exactly what we shall need.

\begin{defn} \label{dfn: supermartingale deflators}
The class of \textsl{equivalent supermartingale deflators} is defined as
\[
\Def_\K := \set{D \geq 0 \such D_0 = 1,\ D_T > 0, \textrm{ and } D
W^\pi \text{ is supermartingale } \forall \pi \in \Pi_\K}.
\]
If there exists an element $D^* \in \Def_\K$ of the form $D^* \equiv 1/W^\rho$ for some $\rho \in
\Pi_\K\,$, we call $D^*$ a \textsl{tradeable supermartingale deflator}.
\end{defn}

If a tradeable supermartingale deflator $D^* \equiv 1/W^\rho$
exists, then the relative wealth process $W^\pi / W^\rho$ is a
supermartingale for all $\pi \in \Pi_\K$, i.e., $\rho$ is the \num
portfolio. Thus, a tradeable supermartingale deflator exists, if and
only if a \num portfolio $\rho$ exists and $W^\rho_T < \infty$,
$\prob$-a.s.; and then it is unique.

An equivalent supermartingale measure $\qprob$ generates an
equivalent supermartingale deflator through the positive martingale
$D_t = \pare{\ud \qprob / \ud \prob} |_{\F_t}$. Then we have
$\QprobC \subseteq \Def_\K$ (for the class $\QprobC$ of equivalent
$\K$-supermartingale measures of Definition \ref{dfn: equivalnet
supermart measure}),  thus $\QprobC \neq \emptyset \Rightarrow
\Def_\K \neq \emptyset$. In general, the elements of $\Def_\K$ are
just supermartingales, not martingales, and the inclusion $\QprobC
\subseteq \Def_\K$ is strict; more importantly, the implication
$\,\Def_\K \neq \emptyset \Rightarrow \QprobC \neq \emptyset\,$ does
\emph{not} hold, as we now show.

\begin{ex}
Consider the the three-dimensional Bessel process Example \ref{ex:
Bessel} on the finite time-horizon $[0,1]$.
Since $\rho = 1$ is the \num portfolio,   $D^* = 1/S$ is
a tradeable supermartingale deflator, so $\,\Def_\K \neq \emptyset\,$.
As we have already seen, NA fails, thus we must have $\,\QprobC =
\emptyset\,$.
\end{ex}

The set $\Def_\K$ of equivalent supermartingale deflators appears as
the range of optimization in the ``dual'' of the utility
maximization problem (\ref{eq: utility optimization}) in
\cite{Kramkov-Schachermayer: Asymptotic elast and util function}. It
has appeared before in some generalization of Kramkov's Optional
Sampling Theorem by Stricker and Yan \cite{Stricker-Yan}, as well as
in Schweizer \cite{Schweizer: mart densities} under the name
``martingale densities'' (in both of these works, $\Def$ consisted
of positive local martingales).

As we shall see soon, \emph{it is the condition $\Def_\K \neq
\emptyset$, rather than $\QprobC \neq \emptyset$, that is needed in
order to solve the utility maximization problem (\ref{eq: utility
optimization})}.

\smallskip

The existence of an equivalent supermartingale deflator has some
consequences for the class of admissible wealth processes.

\begin{prop} \label{prop: if defl exist, wealth proc are semimarts up to T}
If $\,\Def_\K \neq \emptyset\,$, then for every $\pi \in \Pi_\K$ the
wealth process $W^\pi$ is a semimartingale up to time $T$ (for this
concept you can consult Remark \ref{rem: semimarts up to T from
semimart up to inf} in the Appendix). In particular, $\lim_{t \to
\infty} W^\pi_t$ exists on $\{ T = \infty \}$.
\end{prop}

\proof Pick $D \in \Def_\K$ and $\pi \in \Pi_\K$. Since $D W^\pi$ is
a positive supermartingale, Lemma \ref{lem: positive supermart is
semimart to infty} gives that $D W^\pi$ is a semimartingale up to
$T$. Since $D$ is also a positive supermartingale with $D_T > 0$,
$1/D$ is a semimartingale up to $T$, again by Lemma \ref{lem:
positive supermart is semimart to infty}. It follows that $W^\pi =
(1/D) D W^\pi$ is a semimartingale up to $T$. \qed

\medskip

In order to complete the discussion, we mention that if a tradeable
supermartingale deflator $D^*$ exists, Jensen's inequality and the
supermartingale property of $D W^\rho \equiv D / D^*$ for all $D \in
\Def_\K  $ imply $\expec [ - \log D^*_T ] = \inf_{D \in \Def_\K}
\expec [ - \log D_T ]$. This can be viewed as an optimality property
of the tradeable supermartingale deflator, dual to log-optimality of
the \num portfolio as discussed in subsection \ref{subsec: relative
log-optimal}. We can also consider it as a \emph{minimal reverse
relative entropy} property of $D^*$ in the class $\Def_\K$. Let us
explain: for every element $D \in \Def_\K$ that is actually a
uniformly integrable martingale, consider the probability measure
$\qprob$ defined by $\qprob(A) = \expec[D_T \indic_A]$; then, the
quantity $H (\prob \such \qprob) := \expec^\qprob [D_T^{-1} \log
\pare{D_T^{-1}}] = \expec [- \log D_T]$ is the \textsl{relative
entropy} of $\prob$ with respect to $\qprob$. In general, even when
$D$ is not a martingale, we could regard $\expec [- \log D_T ]$ as
the relative entropy of $\prob$ with respect to $D$. The qualifier
``reverse'' comes from the fact that one usually considers
minimizing the entropy of \emph{another} equivalent probability
measure $\qprob$ with respect to the \emph{original} $\prob$
(so-called   \textsl{minimal entropy measure}). For further details
and history we refer the reader to Example 7.1 of Karatzas \& Kou
\cite{Karatzas-Kou}, Schweizer \cite{Schweizer: mmm} where the
minimal reverse relative entropy property of the ``minimal
martingale measure'' for continuous asset-price processes is
discussed, as well as Goll and R\"uschendorf \cite{Goll - Rusch:
minimal} where a general discussion of minimal distance martingale
measures is made (of which the minimal reverse entropy martingale
measure is a special case).

\subsection{The second main result} \label{subsec: second main result}

Here is our second main result, which places the \num portfolio in the
context of arbitrage.

\begin{thm} \label{thm: num iff def-non-empty iff NUPBR}
For a financial model described by the stock-price process $S$ and the predictable closed convex
constraints $\K$, the following are equivalent:
\begin{enumerate}
    \item The \num portfolio exists and $W^\rho_T < \infty$.
    \item The set $\,\Def_\K\,$ of equivalent supermartingale deflators  is
    non-empty.
    \item The \NUPBR condition holds.
\end{enumerate}
\end{thm}

The implication (1) $\Rightarrow$ (2) is trivial: $(W^\rho)^{-1}$ is
an element of $\Def_\K$ (observe that we need $W^\rho_T < \infty$ to
get $(W^\rho_T)^{-1} > 0$ as required in the definition of
$\Def_\K$).

For the implication (2) $\Rightarrow$ (3), start by assuming that
$\Def_\K \neq \emptyset$ and pick $D \in \Def_\K$. We wish to show
that the collection $(W^\pi_T)_{\pi \in \Pi_\K}$, the terminal
values of positive wealth processes with $W^\pi_0 = 1$ is bounded in
probability. Since $D_T
>0$, this is equivalent to showing that the collection $\{D_T W^\pi_T \such \pi \in \Pi_\K\}$
is bounded in probability. But since every process $D W^\pi$ for
$\pi \in \Pi_\K$ is a positive supermartingale we have $\prob [ D_T
W^\pi_T > a ] \leq a^{-1} \expec[D_T W^\pi_T] \leq a^{-1} \expec[D_0
W^\pi_0] = a^{-1}$, for all $a > 0$; this last estimate does not
depend on $\pi \in \Pi_\K$, and we are done.

Implication (3) $\Rightarrow$ (1) is much harder to prove. One has
to analyze what happens when the \num portfolio fails to exist; we
do this in the next subsection.

\medskip

Theorem \ref{thm: num iff def-non-empty iff NUPBR} provides the
equivalent of the FTAP when we only have convex, but not
necessarily conic, constraints. Since the existence of a \num
portfolio $\rho$ with $W_T^\rho < \infty$ is equivalent to
$\Psi_T(B, C, \eta) < \infty$ according to Remark \ref{rem:
deterministic functional deciding on num}, we obtain also a
partial answer to our second question, regarding the
characterization of free lunches in terms of predictable
characteristics from \S \ref{subsubsec: free lunches and pred
char}; the full answer will be given in the next subsection
\ref{subsec: what happens when num fails to exist}. Finally, the
question on utility maximization posed at \S \ref{subsubsec:
connection with util maxim} will be tackled in subsection
\ref{subsec: Utility optimization}.

\begin{rem}
Conditions (2) and (3) of Theorem \ref{thm: num iff def-non-empty
iff NUPBR} remain invariant by an equivalent change of probability
measure. Thus, existence of the \num portfolio remains unaffected
also, although the \num portfolio itself will change. Though a
pretty reasonable conjecture to have made at the outset, this does
not seem to follow directly from the definition of the \num
portfolio.

The above fail if we only consider \emph{absolutely continuous}
changes of measure (unless $S$ is continuous). One would guess that
NUPBR should hold, but non-equivalent changes of probability measure
might enlarge the class of admissible wealth processes, since now
the positivity condition for wealth processes is easier satisfied
--- in effect, the natural constraint set $\K_0$ can be larger.
Consider, for example, a finite time-horizon case where, under
$\prob$, $X$ is a driftless compound Poisson process and $\{-1/2,
1/2\}$ is exactly the support of $\nu$. Here, $\K_0 = [-2, 2]$ and
$X$ itself is a martingale. Now, consider the simple absolutely
continuous change of measure that transforms the jump measure to
$\nu_1 (\ud x) := \indic_{\{x > 0 \}} \nu (\ud x)$; then, $\K_0 =
[-2, \infty)$ and of course NUIP fails.
\end{rem}

\begin{rem}
Theorem \ref{thm: num iff def-non-empty iff NUPBR} together with
Proposition \ref{prop: if defl exist, wealth proc are semimarts up
to T} imply that under \NUPBRC all wealth processes $W^\pi$ for $\pi
\in \Pi_\K$ are semimartingales up to infinity. Thus, under \NUPBRC
the assumption about existence of $\lim_{t \to \infty} W^\pi_t$ on
$\{T = \infty\}$ needed for the NA, and the NFLVR conditions in
Definition \ref{dfn: arbitrage notions} is superfluous.
\end{rem}

\subsection{Consequences of non-existence of the \num portfolio} \label{subsec: what happens when num fails to exist}

In order to finish the proof of Theorem \ref{thm: num iff
def-non-empty iff NUPBR}, we need to describe   \emph{what goes
wrong when the \num portfolio fails to exist}. This can happen in
two ways. First, the set $\{\I \cap \check{\K} \neq \emptyset\}$ may
not have zero $\prob \otimes G$-measure; in this case, Proposition
\ref{prop: pred charact of (NUIP)} shows that one can construct an
unbounded increasing profit, the most egregious form of arbitrage.
Secondly, when   $(\prob \otimes G) (\{\I \cap \check{\K}  \neq
\emptyset\})=0$, the constructed predictable process $\rho$ can fail
to be $X$-integrable (up to time $T$). The next definition prepares
the ground for Proposition \ref{prop: with num you can never go
wrong}, which describes what happens in this latter case.

\begin{defn} \label{dfn: limsup in prob}
Consider a sequence $(f_n)_{n \in \Natural}$ of random variables. Its \textsl{superior limit in
the probability sense}, $\prob$-$\limsup_{n \to \infty} f_n$, is defined as the essential infimum
of the collection $\set{g \in \F \such \lim_{n \to \infty} \prob[f_n \leq g] = 1}$.
\end{defn}

It is obvious that the sequence $(f_n)_{n \in \Natural}$ of random variables is unbounded in
probability if and only if $\prob$-$\limsup_{n \to \infty} \abs{f_n} = + \infty$ with positive
probability.

\begin{prop} \label{prop: with num you can never go wrong}
Assume that the predictable set $\{ \I \cap \check{\K} \neq\emptyset
\}$ has zero $\prob \otimes G$-measure, and let $\rho$ be the
predictable process constructed in Theorem \ref{thm: existence of
the num for semimarts}. Pick any sequence $(\theta_n)_{n \in
\Natural}$ of $[0,1]$-valued predictable processes with $\lim_{n \to
\infty} \theta_n = \indic$ holding $\prob \otimes G$-almost
everywhere, such that $\rho_n := \theta_n \rho$ has bounded support
and is $X$-integrable for all $n \in \Natural$. Then
$\overline{W}^\rho_T := \prob$-$\limsup_{n \to \infty} W^{\rho_n}_T$
is a $(0, + \infty]$-valued random variable, and does not depend on
the choice of the sequence $(\theta_n)_{n \in \Natural}$. On $\{
(\psi^\rho \cdot G)_T < + \infty \}$, the random variable
$\overline{W}^\rho_T$ is an actual limit in probability and
\[
\{ \overline{W}^\rho_T = + \infty \} = \set{(\psi^\rho \cdot G)_T =
+ \infty};
\]
in particular, $\prob [\,\overline{W}^\rho_T  = + \infty\,]
> 0$ if and only if $\rho$ fails to be $X$-integrable up to $T$.
\end{prop}

The above result says, in effect, that  {\sl closely following a
\num portfolio which is not $X$-integrable up to time $T$, one can
make arbitrarily large gains with fixed, positive probability}.
There are many ways to choose the sequence $(\theta_n)_{n \in
\Natural}$; a particular example is $\theta_n := \indic_{\Sigma_n}$
with $\Sigma_n := \set{(\omega,t) \in \dbra{0, T \wedge n} \such
|\rho(\omega,t)| \leq n }$.

Proposition \ref{prop: with num you can never go wrong} is proved in
section \ref{sec: proof that with num you can never go wrong}; it
answers in a definitive way the question regarding the description
of free lunches in terms of predictable characteristics, raised in
\S \ref{subsubsec: free lunches and pred char}: When \NUPBRC fails
(equivalently, when the \num portfolio fails to exist, or exists but
$\prob[W^\rho_T = \infty] > 0$), there is a way to \emph{construct}
the unbounded profit with bounded risk (UPBR) using knowledge of the
triplet of predictable characteristics.

\smallskip
\noindent \emph{Proof of Theorem \ref{thm: num iff def-non-empty iff
NUPBR}:} Assuming Proposition \ref{prop: with num you can never go
wrong}, we are   now in a position to show the implication (3)
$\Rightarrow$ (1) of Theorem \ref{thm: num iff def-non-empty iff
NUPBR} and complete its proof. Suppose then that the \num portfolio
fails to exist. Then, we either we have opportunities for unbounded
increasing profit, in which case NUPBR certainly fails; or the
predictable process $\rho$ of Theorem \ref{thm: existence of the num
for semimarts} exists but is not $X$-integrable up to time $T$, in
which case Proposition \ref{prop: with num you can never go wrong}
provides an UPBR. \qed

\begin{rem} In the context of Proposition \ref{prop:
with num you can never go wrong}, suppose that $\{ \I \cap
\check{\K} \neq\emptyset \}$ has zero $\prob \otimes G$-measure. The
failure of $\rho$ to be $X$-integrable up to time $T$ can happen in
two ways. Start by defining $\tau := \inf \{ t \in [0, T] \such
\pare{\psi^\rho \cdot G}_t = + \infty \}$ and $\,\tau_n := \inf \{ t
\in [0, T] \such (\psi^\rho \cdot G)_t \geq n \}\,,~n \in \Natural
\,$. We consider two cases.

First,  suppose  $\tau > 0$ and $(\psi^\rho \cdot G)_\tau = +
\infty$; then $\tau_n < \tau$ for all $n \in \Natural$ and $\tau_n
\uparrow \tau$. By using the sequence $\rho_n := \rho \,
\indic_{\dbra{0, \tau_n}}$ it is easy to see that $\lim_{n \to
\infty} W^{\rho_n}_\tau = + \infty$ almost surely --- this is
because $\{ (W_t^\rho)^{-1},  \, 0 \leq t < \tau \}$ is a
supermartingale. An example where this happens in finite time is
when the returns process $X$ satisfies $\ud X_t = (1 - t)^{-1/2} \ud
t + \ud \beta_t$, where $\beta$ is a standard one-dimensional
Brownian motion. Then $\,\rho_t = (1 - t)^{-1/2}\,$ and thus
$\,\pare{\psi^\rho \cdot G}_t = \int_0^t (1 - u)^{-1} \ud u\,$,
which   gives $\,\tau \equiv 1\,$.

\smallskip
With the notation set-up above, let us now give an example with
$(\psi^\rho \cdot G)_\tau < + \infty$. Actually, we shall only
time-reverse the example we gave before and show that in this case
$\tau \equiv 0$. To wit, take the stock-returns process now to be
$\,\ud X_t = t^{-1/2} \ud t + \ud \beta_t\,$; then  $\rho_t =
t^{-1/2}$ and $\,( \psi^\rho \cdot G )_t = \int_0^t u^{-1} \ud u = +
\infty\,$ for all $\,t  > 0\,$ so that $\tau = 0$. In this case we
cannot invest in $\rho$ as before in a ``forward'' manner, because
it has a ``singularity'' at $t=0$ and we cannot take full advantage
of it. This is basically what makes the proof of Proposition
\ref{prop: with num you can never go wrong} non-trivial.

In the case of a continuous-path semimartingale $X$ without
portfolio constraints (as the one described in this example),
Delbaen \& Schachermayer \cite{DS: FTAP absolutely continuous} and
Levental \& Skorohod \cite{Lev-Skor: absence of arbitrage} show that
one can actually create ``instant arbitrage'', i.e., a non-constant
wealth process that never falls below its initial capital    (almost
the definition of an increasing unbounded profit, but weaker, since
the wealth process is not assumed to be increasing). In the presence
of jumps, it is an open question whether one can still construct
this instant arbitrage --- we could not. \qquad $\diamond$ \end{rem}

\subsection{Application to Utility Optimization} \label{subsec: Utility optimization}

Here we tackle the question that we raised in \S \ref{subsubsec:
connection with util maxim}. We show that NUPBR  \emph{is the
minimal condition that allows one to solve the utility maximization
problem (\ref{eq: utility optimization}).  }

\begin{rem} \label{rem: on finiteness of value functions of utility}
The optimization problem (\ref{eq: utility optimization}) makes
sense only if $u(w) < \infty$. Since $U$ is concave, if $u(w) < +
\infty$ for \emph{some} $w >0$, then $u(w) < + \infty$ for
\emph{all} $w > 0$ and $u$ is continuous, concave and increasing.
When   $u(w) = \infty$ holds for some (equivalently, all) $w > 0$,
there are two cases. Either the supremum in (\ref{eq: utility
optimization}) is not attained, so there is no solution; or, in case
there exists a portfolio with infinite expected utility, concavity
of $U$ implies that there will be infinitely many of them.
\end{rem}

We begin with the  {\sl negative} result: when \NUPBRC fails,
(\ref{eq: utility optimization}) cannot be solved.

\begin{prop} \label{prop: if NUPBR fails, no utility optimization}
Assume that \emph{NUPBR$_\K$} fails. Then, for any utility function
$U$, the corresponding utility maximization problem either does not
have a solution, or has infinitely many.

\noindent
 More precisely: If $U(\infty) = + \infty$, then $u(w) =
+ \infty$ for all $w > 0$, so we either have no solution (when the
supremum is not attained) or infinitely many of them (when the
supremum is attained); whereas if $\,U(\infty) < + \infty\,$, there
is no solution.
\end{prop}

\proof Since \NUPBRC fails, pick an $\epsilon > 0$ and a sequence
$(\pi_n)_{n \in \Natural}$ of elements of $\Pi_\K$ such that, with
$A_n := \set{W^{\pi_n}_T \geq n}$, we have $\prob[A_n] \geq
\epsilon$ for all $n \in \Natural$.

If   $U(\infty) = + \infty$, then it is obvious that, for all $w >
0$ and $n \in \Natural$, we have $u(w) \geq \expec[U(w W^{\pi_n}_T)]
\geq \epsilon U(w n)$;   thus $u(w) = + \infty$ and we obtain the
result stated in the proposition in view of Remark \ref{rem: on
finiteness of value functions of utility}.

Now  suppose   $U(\infty) < \infty$; then $U(w) \leq u(w) \leq
U(\infty) < \infty$ for all $w > 0$. Furthermore, $u$ is also
concave, thus continuous. Pick any $w > 0$, suppose that $\pi \in
\Pi_\K$ is optimal for $U$ with initial capital $w$, and observe:
$u(w + n^{-1}) \geq \expec [U(w W^{\pi}_T + n^{-1} W^{\pi_n}_T)]
\geq \expec [U(w W^\pi_T + \indic_{A_n})]$, as well as
\[
U(w W^\pi_T + \indic_{A_n}) \ = \ U(w W^\pi_T) \, \indic_{\Omega
\setminus A_n} + U(w W^\pi_T + 1) \, \indic_{A_n}\,.
\]
Pick $M>0$ large enough so that $\prob[w W^{\pi}_T \leq M] \geq 1 -
\epsilon/2$; then, for $\, 0 <y \le M$ the concavity of $U$ gives
$\,U(y+1) - U(y) \geq U(M+1) - U(M)=:b\, $. Therefore,
\[
U(w W^\pi_T + 1) \ \geq \ \big(U(w W^\pi_T) + b\big)\, \indic_{\{w
W^{\pi}_T \leq M\}}\, +\, U(w W^\pi_T)\, \indic_{\{w W^{\pi}_T >
M\}}\,.
\]
Combining the two previous estimates, we get
\[
U(w W^\pi_T + \indic_{A_n}) \ \geq \ U(w W^\pi_T)\, +\, b\,
\indic_{A_n \cap \{w W^{\pi}_T \leq M\}}\,.
\]
Since $\prob[A_n] \geq \epsilon\,$  we get $\,\prob[A_n \cap \{w
W^{\pi}_T \leq M\}] \geq \epsilon/2\,$, and setting $a := b
\epsilon/2$  we obtain $u(w + n^{-1}) \geq \expec[U(w W^\pi_T +
\indic_{A_n})] \geq \expec[U(w W^\pi_T)] + a = u(w) + a$ for all $n
\in \mathbb{N}$ which contradicts the continuity of $u(\cdot)\,.
\qquad $ \qed

\medskip

Having discussed what happens when \NUPBRC fails, let us now assume
that it holds. We shall  assume a little   more structure on the
utility function under consideration, namely, that it is
continuously differentiable and satisfies the \textsl{Inada
conditions} $U'(0) = +\infty$ and $U'(+\infty) = 0$.

The \NUPBRC condition is equivalent to the existence of a \num
portfolio $\rho$. Since all wealth processes  become
supermartingales when divided by $W^\rho$, we conclude that the
change of \num that utilizes $W^\rho$ as a benchmark produces a
market for which the \emph{original}  $\prob$ is a supermartingale
measure (see Delbaen and Schachermayer \cite{DS: numeraire} for this
``change of num\'eraire'' technique). In particular, \NFLVRC holds
and the ``optional decomposition under convex constraints'' results
of \cite{FoKr} allow us to write down the \emph{superhedging
duality}
\[
\inf \set{w > 0 \such \exists \ \pi \in \Pi_\K \text{ with } w
W^{\pi}_T \geq H} = \sup_{D \in \Def_\K} \expec [D_T H]\,,
\]
valid for any positive, $\F_T$-measurable random variable $H$. This
``bipolar'' relationship then implies that the utility optimization
problem admits a solution (when its value is finite). We send the
reader to the papers \cite{Kramkov-Schachermayer: Asymptotic elast
and util function, Kramkov-Schachermayer: nec and suf incomplete
markets} for more information.

\subsection{A word on the additive model} \label{subsec: on additive model}

All the results stated up to now hold also when the stock-price
processes $S^i$ are \emph{not necessarily positive} semimartingales.
Indeed, suppose that we start with initial prices $S_0$, introduce
$Y := S - S_0$, and define the admissible (discounted) wealth
processes class to be generated by strategies $\theta \in
\Pre(\Real^d)$ via $W = 1 + \theta \cdot S = 1 + \theta \cdot Y$,
where we force $W > 0$, $W_- > 0$. Here, $\theta$ is the
\emph{number} of shares of stocks in our portfolio. Then, with $\pi
:= (1 / W_-) \theta$, it follows that we can write $W = \Exp(\pi
\cdot Y)$. We do not necessarily have $\Delta Y > -1$ anymore, but
this was never used anywhere; the important thing is that
admissibility implies $\pi^\top \Delta Y > -1$. Observe that now
$\pi$ does not have a nice interpretation as it had in the case of
the multiplicative model.

A final note on constraints. One choice is to require $\theta \in
W_- \K$, which is completely equivalent to $\pi \in \K$. A more
natural choice would be to enforce them on investment proportions,
i.e., to require $(\theta^i S^i_- / W_-)_{1 \leq i \leq d} \in \K$,
in which case we get $\pi \in \widehat{\K}$, where $\widehat{\K} :=
\{ x \in \Real^d \such (x^i S^i_-)_{1 \leq i \leq d} \in \K \}$ is
predictable.

\smallskip

\section{Proof of Proposition \ref{prop: pred charact of (NUIP)}
 \label{sec: The
NUIP condition}  on the NUIP Condition}

\subsection{If $\{\I \cap \check{\K} \neq \emptyset\}$ is $\prob \otimes G$-null, then NUIP holds}
\label{subsec: (NIP) implies I cap K is empty}

Let us suppose that $\pi$ is a portfolio with unbounded increasing
profit; we shall show that $\{\I \cap \check{\K} \neq \emptyset\}$
is not $\prob \otimes G$-null. By   definition  then $\{\pi \in
\check{\K}\}$ has full $\prob \otimes G$-measure, so we wish to
prove that $\{\pi \in \I\}$ has strictly positive $\prob \otimes
G$-measure.

Now  $W^\pi$ has to be a non-decreasing process, which means that
the same   holds for $\pi \cdot X$. We also   have $\pi \cdot X \neq
0$ with positive probability. This means that the predictable set
$\{\pi \notin \N\}$ has strictly positive $\prob \otimes G$-measure,
and it will suffice to show that properties (1)--(3) of Definition
\ref{dfn: instant_arb_opport} hold $\prob \otimes G$-a.e.

Because $\pi \cdot X$ is increasing, we get $\indic_{\{\pi^\top x <
0\}} * \mu = 0$, so that $\nu [\pi^\top x < 0] = 0$, $\prob \otimes
G$-a.e. In particular, $\pi \cdot X$ is of finite variation, so we
must have   $\pi \cdot X^\co = 0$, and this translates into
$\pi^\top c = 0$, $\prob \otimes G$-a.e. For the same reason, one
can decompose
\begin{equation} \label{eq: decomp of UIP}
\pi \cdot X \,= \, \left( \pi \cdot B - [\pi^\top \hx] * \eta
\right) + [\pi^\top x]
* \mu\,.
\end{equation}
The last term $[\pi^\top x] * \mu$ in this decomposition is a
pure-jump increasing process, while for the sum of the terms in
parentheses we have from (\ref{eq: when G jumps}):
\[
\Delta \Big( \pi \cdot B - [\pi^\top \hx] * \eta \Big) = \Big(
\pi^\top b - \int \pi^\top \hx \nu (\ud x) \Big) \Delta G = 0\,.
\]
It follows that the   term in parentheses on the right-hand side
of equation (\ref{eq: decomp of UIP}) is the continuous part of
$\pi \cdot X$ (when seen as a finite variation process) and thus
has to be increasing. This translates into the requirement
$\pi^\top b - \int \pi^\top \hx \nu (\ud x) \geq 0$, $\prob
\otimes G$-a.e., and ends the proof.

\subsection{The set-valued process $\I$ is predictable}
In proving the other half of Proposition \ref{prop: pred charact of
(NUIP)}, we need to select a predictable process from the set $\{\I
\cap \check{\K} \neq \emptyset\}$. For this, we shall have to prove
that $\I$ is a predictable set-valued process; however, $\I$ is not
closed, and closedness of sets is crucial when trying to apply
measurable selection results. For this reason we have to go through
some technicalities first.

Given a   triplet $(b,c,\nu)$ of predictable characteristics and
$a>0$, define $\I^a$ to be the set-valued process such that (1)--(3)
of Definition \ref{dfn: instant_arb_opport} hold, as well as
\begin{equation} \label{eq: further requirement for I^a}
\xi^\top b + \int \frac{\xi^\top x}{1 + \xi^\top x} \indic_{\{ |x| >
1\}} \nu (\ud x) \, \geq \frac{1}{a}.
\end{equation}
The following lemma sets forth properties of these sets that we
shall find useful.

\begin{lem} \label{lem: approx of immed arb sets and (NUIP) holds for original iff holds for trunc}
With the previous definition we have:
\begin{enumerate}
    \item $\I^a$ is increasing in $a > 0$; we have $\I^a \subseteq
    \I$ and   $\I =  \bigcup_{a > 0} \I^a$. In particular, $\I \cap
    \check{\K}
    \neq \emptyset$ if and only if
    $\I^a \cap \check{\K} \neq \emptyset$ for all large enough $a > 0$.
    \item For all $a > 0$, $\I^a$ takes values in closed and convex subsets of $\Real^d$.
\end{enumerate}
\end{lem}

\proof In the course of the proof, we suppress dependence of
quantities on $(\omega, t)$.

Because of conditions (1)--(3) of Definition \ref{dfn:
instant_arb_opport},   the left-hand-side of (\ref{eq: further
requirement for I^a}) is well-defined (the integrand is positive
since $\nu[\xi^\top x < 0] = 0$) and has to be positive. In fact,
for $\xi \in \I$, it has to be \emph{strictly} positive, otherwise
$\xi \in \N$. The fact that $\I^a$ is increasing for $a > 0$ is
trivial, and part (1) of this lemma follows.

For part (2), we show first that $\I^a$ is closed. Observe that the
set $\{ \xi \in \Real^d \such \xi^\top c = 0 \textrm{ and } \nu [
\xi^\top x < 0 ] = 0 \}$ is closed in $\Real^d$. For $\xi$ on this
last set, $x \mapsto \xi^\top x$ is non-negative for all $x \in
\Real^d$ on a set of full $\nu$-measure. For a sequence $(\xi_n)_{n
\in \Natural}$ in $\I^a$ with $\lim_{n \to \infty} \xi_n = \xi$,
Fatou's lemma gives
\[
\int \xi^\top \hx \nu (\ud x) \leq \liminf_{n \to \infty} \int
\xi_n^\top \hx \nu (\ud x) \leq \liminf_{n \to \infty} \left(
\xi_n^\top b \right)\,=\, \xi^\top b\,,
\]
so that $\xi$ satisfies   (3) of Definition \ref{dfn:
instant_arb_opport} also. The measure $\indic_{\{|x| > 1\}} \nu(\ud
x)$ (the ``large jumps'' part of the L\'evy measure $\nu$) is
finite, and bounded convergence gives
\[
\xi^\top b + \int \frac{\xi^\top x}{1 + \xi^\top x} \indic_{\{ |x| \geq 1\}} \nu (\ud x) = \lim_{n
\to \infty} \Big\{ \xi_n^\top b + \int \frac{\xi_n^\top x}{1 + \xi_n^\top x} \indic_{\{ |x| \geq 1
\}} \nu (\ud x) \Big\} \geq a^{-1}\,.
\]
This establishes that $\I^a$ is closed. Convexity follows from the
fact that the function $x \mapsto x/(1+x)$ is concave on $(0,
\infty)$. \qed

\medskip

In view of $\,\I = \bigcup_{n \in \Natural} \I^n$ and Lemma
\ref{lem: union and intersects of measur set-valued processes}, in
order to prove predictability of $\I$ we only have to prove
predictability of $\,\I^a$. To this end, we define the following
real-valued functions, with arguments in $(\Omega \times \Real_+)
\times \Real^d$ (once again, suppressing their dependence on
$(\omega,t) \in \dbra{0,T}$):
\begin{eqnarray*}
  z_1 (\p) &=& \p^\top c, \ \ \ \ \ z_2 (\p) \ \ = \ \ \int \frac{((\p^\top x)^-)^2}{1 + ((\p^\top x)^-)^2}
\nu
(\ud x), \\
  z_3^n (\p) &=& \p^\top b - \int \p^\top x \indic_{\{n^{-1} < |x| \leq 1\}} \nu
(\ud x),
  \text{ for all } n \in \Natural, \text{ and}\\
  z_4 (\p) &=& \p^\top b + \int \frac{\p^\top x}{1 + \p^\top x} \indic_{\{ |x| > 1\}} \nu
(\ud x).
\end{eqnarray*}
Observe that all these functions are predictably measurable in
$(\omega, t) \in \Omega \times \Real_+$ and continuous in $\p$
(follows from applications of the dominated convergence theorem). In
a limiting sense, consider formally $z_3(\p) \equiv z_3^\infty (\p)
= \p^\top b - \int \p^\top \hx \nu (\ud x)$;   observe though that
this function might not   be well-defined: both the positive and
negative parts of the integrand might have infinite $\nu$-integral.
Consider also the sequence $\Af^a_n := \{ \p \in \Real^d \such
z_1(\p) = 0,\ z_2(\p) = 0,\ z_3^n (\p) \geq 0,\ z_4 (\p) \geq a^{-1}
\}$ of set-valued processes for $n \in \Natural$, of which the
``infinite'' version coincides with $\,\I^a\,$: $\I^a \equiv
\Af^a_\infty := \{\p \in \Real^d \such z_1(\p) = 0,\ z_2(\p) = 0,\
z_3 (\p) \geq 0,\ z_4 (\p) \geq a^{-1} \}$. Because $z_2(\p) = 0$,
the function $z_3$ is well-defined (though not necessarily finite,
since it can equal $- \infty$). In any case, for any   $\p$ with $\,
z_2(\p) = 0 \,$  we have  $\downarrow \lim_{n \to \infty} z_3^n (\p)
= z_3 (\p)$; so   the sequence $(\Af^a_n)_{n \in \Natural}$ is
decreasing, and   $\downarrow \lim_{n \to \infty} \Af^a_n = \I^a$.
But each $\Af^a_n$ is closed and predictable (refer to Lemmata
\ref{lem: union and intersects of measur set-valued processes} and
\ref{lem: Caratheodory makes measurable sets}), and thus so is
$\,\I^a\,$.

\begin{rem} \label{rem: predictability of I cap K neq emptyset}
Since $\{\I \cap \check{\K} \neq \emptyset\} = \bigcup_{n \in
\Natural} \{\I^n \cap \check{\K} \neq \emptyset\}$ and the random
set-valued processes $\I^n$ and $\check{\K}$ are closed and
predictable, Appendix A shows that the set $\{ \I \cap \check{\K}
\neq \emptyset \}$ is predictable.
\end{rem}

\subsection{NUIP implies that $\{\I \cap \check{\K}
\neq \emptyset \}$ is $\prob \otimes G$-null}

We are now ready to finish the proof of Proposition \ref{prop: pred charact of (NUIP)}. Let us
suppose that $\{\I \cap \check{\K} \neq \emptyset\}$ is not $\prob \otimes G$-null; we shall
construct an unbounded increasing profit.

Since $\I = \bigcup_{n \in \Natural} ( \{ \p \in \Real^d \such |\p|
\leq n \} \cap \I^n)$, where $\I^n$ is the set-valued process of
Lemma \ref{lem: approx of immed arb sets and (NUIP) holds for
original iff holds for trunc}, there exists $n \in \Natural$ such
that the \emph{convex, closed} and \emph{ predictable} set-valued
process $\,\B^n := \{\p \in \Real^d \such | \p | \leq n \} \cap \I^n
\cap \check{\K}\,$ has $\,(\prob \otimes G) (\{\B^n \neq
\emptyset\}) > 0\,$. From Theorem \ref{thm: measurable maximum},
there exists a predictable process $\pi$ such that $\pi(\omega, t)
\in \B^n(\omega, t)$ when $\,\B^n(\omega, t) \neq \emptyset\,$, and
$\,\pi(\omega, t)=0\,$ if $\,\B^n(\omega, t) = \emptyset\,$. This
$\pi$ is bounded, so $\pi \in \Pi_\K$. The reasoning of subsection
\ref{subsec: (NIP) implies I cap K is empty}, now ``in reverse",
gives that $\pi \cdot X$ is non-decreasing;   the same is then true
of $W^\pi$. Thus, we \emph{must} have $\prob[W^\pi_\infty
> 1] > 0$, otherwise   $\pi
\cdot X \equiv 0$, which is impossible since $(\prob \otimes G) (\{ \pi \notin \N \}) > 0$ by
construction.

\smallskip

\section{Proof of the Main Theorem \ref{thm: existence of the num for
semimarts}} \label{sec: The Numeraire for General Semimartingales}

We saw in Lemma \ref{lem: necess and suff for numeraire} that if the
\num portfolio $\rho$ exists, it has to satisfy $\rel(\pi \such
\rho) \leq 0$ pointwise, $\prob \otimes G$-a.e. In order to find
necessary and sufficient conditions for the existence of a
(predictable) process $\rho$ that satisfies this inequality, it
makes sense first to consider the corresponding static,
deterministic problem.

\subsection{The Exponential L\'evy market
case} \label{subsec: The Numeraire for Exp Levy}

L\'evy processes correspond to constant, deterministic triplets of
characteristics with respect to the natural time flow $G(t) = t$, so
we shall take in this subsection $X$ to be a L\'evy process with
deterministic \textsl{L\'evy triplet} $(b,c,\nu)$; this means $B_t =
b t$, $C_t = c t$ and $\eta(\ud t, \ud x) = \nu (\ud x) \ud t$ in
the notation of subsection \ref{subsec: stock-price model}. We also
take $\K$ to be a closed convex subset of $\Real^d$; recall that $\K
\subseteq \K_0$, where $\K_0 := \{\pi \in \Real^d \such \nu
[\pi^\top x< -1] = 0\}$.

The following result is the deterministic analogue of Theorem
\ref{thm: existence of the num for semimarts}.

\begin{lem} \label{thm: deterministic_solution}
Let $(b,c,\nu)$ be a L\'evy triplet and $\K$ a closed convex subset of $\Real^d$. Then the
following are equivalent:
\begin{enumerate}
    \item $\I \cap \check{\K} = \emptyset$.
    \item There exists a unique vector $\rho \in \K \cap \N^\bot$ with
    $\nu [ \rho^\top x \leq -1 ] = 0$ such that
    $\,\rel(\pi \such \rho) \leq 0$ holds for all $\pi \in \K$.
    \end{enumerate}
    If the L\'evy measure $\nu$ integrates the logarithm,
    the vector $\rho$ is given as $
    \rho = \arg \max_{\pi \in \K \cap \N^\bot} \g (\pi)$.
    In general, $\rho$ is the limit of
    the optimizers of a sequence of problems, in which $\nu$ is replaced by a sequence of
    approximating measures.
\end{lem}

We have already shown that if (1) fails, then (2) fails as well
(actually, we have argued it for the general semimartingale case;
see Remark \ref{rem: if I cap K neq 0, no solution for rel}). The
proof of the implication (1) $\Rightarrow$ (2) is quite long
--- it can be found in Kardaras \cite{K: arbitrage for Levy}, section
4, where free lunches for exponential L\'evy models are studied in
detail.

\subsection{Integrability of the \num portfolio}

We are close to the proof of our main result. We start with a
characterization of $X$-integrability that the predictable process
$\rho$, our candidate for   \num portfolio, must satisfy. The
following general result is proved in \cite{Cherny}.

\begin{thm} \label{thm: pred charact of integr up to infty, general}
Let $X$ be a $d$-dimensional semimartingale with triplet
 of predictable characteristics is $(b,c,\nu)$,
relative to the canonical truncation function and some operational
clock $G$. A process $\rho \in \Pre(\Real^d)$ is $X$-integrable, if
and only if $(|\widehat{\psi}_i^\rho| \cdot G)_t < \infty$,
$i=1,2,3$, for all $t \in \dbra{0, T}$ holds for the predictable
processes $\, \widehat{\psi}_1^\rho :=\rho^\top c \rho\,$,
\[
  \widehat{\psi}_2^\rho := \int \big( 1 \wedge | \rho^\top x |^2\big)\, \nu (\ud
  x), \ \   \text{ and }~~~
  \ \widehat{\psi}_3^\rho := \rho^\top b + \int \rho^\top x \,
     \big( \indic_{\{ |x| > 1\}} - \indic_{\{ |\rho^\top x| >
    1 \} }\big)\, \nu (\ud x)\,.
\]

\end{thm}

The process $\widehat{\psi}_1^\rho$ controls the quadratic variation
of the continuous martingale part of $\rho \cdot X$; the process
$\widehat{\psi}_2^\rho$ controls the quadratic variation of the
``small-jump'' purely discontinuous martingale part of $\rho \cdot
X$ and the intensity of the ``large jumps''; whereas
$\widehat{\psi}_3^\rho$ controls the drift term of $\rho \cdot X$
when the large jumps are subtracted (it is actually the drift rate
of the bounded-jump part). We use Theorem \ref{thm: pred charact of
integr up to infty, general} to prove Lemma \ref{lem: pred charact
of integr up to infty, special} below, which provides a necessary
and sufficient condition for $X$-integrability of the candidate for
\num portfolio.

\begin{lem} \label{lem: pred charact of integr up to infty, special}
Suppose that $\rho$ is a predictable process with $\nu [\rho^\top
x \leq -1] = 0$ and $\rel(0 \such \rho) \leq 0$. Then  $\rho$ is
$X$-integrable, if and only if the condition $(\psi^\rho \cdot
G)_t (\omega) < \infty$, for all $(\omega, t) \in \dbra{0, T}\,$,
holds for the increasing, predictable process
\[
\psi^\rho \,:= \, \nu[\rho^\top x > 1] + \abs{ \rho^\top b + \int
\rho^\top x \,\big( \indic_{\{ |x| > 1\}} - \indic_{\{ |\rho^\top
x|
> 1 \}} \big)\, \nu (\ud x)}\,.
\]
\end{lem}

\proof We have to show that $G$-integrability of the positive
processes $\psi_1^\rho$ and $|\psi_2^\rho|$ (that add up to
$\psi^\rho$) of \eqref{eq: phi1 and phi2} is necessary and
sufficient for $G$-integrability of the three processes
$\widehat{\psi}_i^\rho$, $i=1,2,3$ of Theorem \ref{thm: pred charact
of integr up to infty, general}. According to this last Theorem,
only the sufficiency has to be proved, since the necessity holds
trivially (recall $\nu [\rho^\top x \leq -1] = 0$). Furthermore,
from the same theorem, the sufficiency will be established if we can
prove that the predictable processes $\widehat{\psi}_1^\rho $ and
$\widehat{\psi}_2^\rho$ are $G$-integrable (note that
$\widehat{\psi}^\rho_3$  is already covered by $\psi^\rho_2$).

Dropping the ``$\rho$'' superscripts, we embark on proving the
$G$-integrability of $\widehat{\psi}_1$ and $\widehat{\psi}_2$,
assuming the $G$-integrability of $\psi_1$ and $\psi_2$ in
(\ref{eq: phi1 and phi2}). The process $\widehat{\psi}_2$ will
certainly be $G$-integrable, if one can show that the positive
process
\[
\widetilde{\psi}_2 \, := \, \int \frac{(\rho^\top x)^2}{1 +
\rho^\top x}\, \indic_{\{|\rho^\top x| \leq 1 \}} \nu (\ud x)\,
+\, \int \frac{\rho^\top x}{1 + \rho^\top x}\, \indic_{\{\rho^\top
x
> 1 \}} \nu (\ud x)
\]
is $G$-integrable. Since both $\,-\rel(0 \such \rho)$ and
$\,\widehat{\psi}_1\,$ are positive processes, we get that
$\widehat{\psi}_1$ and $\widehat{\psi}_2$ will certainly be
$G$-integrable, if we can show that $\,\widehat{\psi}_1 +
\widetilde{\psi}_2 - \rel(0 \such \rho)\,$ is $G$-integrable. But
this last sum is equal to
\[
\rho^\top b \,+ \int \rho^\top x \,\big(\indic_{\{ |x| > 1 \}} -
\indic_{\{| \rho^\top x|
> 1\}}\big)\, \nu (\ud x) \,+\, 2 \int \frac{\rho^\top x}{1 +
\rho^\top x} \,\indic_{\{ \rho^\top x > 1 \}} \nu (\ud x)\,;
\]
the sum of the first two terms equals   $\,\psi_2$, which is
$G$-integrable, and the last (third) term is $G$-integrable because
$\, \psi_1 = \nu [ \rho^\top x >1]\, $ is. \qed

\medskip

In the context of Lemma \ref{lem: pred charact of integr up to
infty, special}, if we wish $\rho$ to be $X $-integrable up to $T$
and not simply $X $-integrable, we have to impose $\psi^\rho_T <
\infty$. This follows from the equivalent characterization of
$X$-integrability up to $T$ in Theorem \ref{thm: pred charact of
integr up to infty, general}, proved in \cite{Cherny}.

Theorem \ref{thm: pred charact of integr up to infty, general}
should be contrasted with Lemma \ref{lem: pred charact of integr up
to infty, special}, where one does not have to worry about the large
negative jumps of $\rho \cdot X$, about the quadratic variation of
its continuous martingale part, or about the quadratic variation of
its small-jump purely discontinuous parts. This follows exactly
because in Lemma \ref{lem: pred charact of integr up to infty,
special} we assume $\nu [\rho^\top x \leq -1] = 0$ and $\rel(0 \such
\rho) \leq 0$: there are not many negative jumps (none above unit
magnitude), and the drift dominates the quadratic variation.

\subsection{Proof of Theorem \ref{thm: existence of the num for semimarts}}

The fact that $\{ \I \cap \check{\K} \neq \emptyset \}$ is
predictable has been shown in Remark \ref{rem: predictability of I
cap K neq emptyset}. The claim (2) follows directly from Lemmata
\ref{lem: necess and suff for numeraire} and \ref{lem: pred charact
of integr up to infty, special}.

For the claims (1.i)--(1.iii), suppose that $\{ \I \cap \check{\K}
\neq \emptyset \}$ has zero $\prob \otimes G$-measure. Set $\Lambda
:= \{ \int\log(1 + |x|) \indic_{\{ |x| > 1 \}}  \nu (\ud x) < \infty
\}$ --- on the predictable set $\Lambda$, the random measure $\nu$
integrates the log. For all $(\omega, t) \in \{ \I \cap \check{\K} =
\emptyset \} \cap \Lambda$, according to Lemma \ref{thm:
deterministic_solution}, there exists a (uniquely defined)
$\rho(\omega, t) \in \Real^d$ with $\rho(\omega, t)^\top \Delta
X(\omega, t)
> -1$ that satisfies $\rel(\pi \such \rho) \leq 0$, and $\g(\rho) =
\max_{\pi \in \K \cap \N^\bot} \g (\pi)$. We also set $\rho = 0$ on
the ($\prob \otimes G$-null) set $\{ \I \cap \check{\K} = \emptyset
\}$.

If $\{ \I \cap \check{\K} = \emptyset \} \cap \Lambda$ has full
$\prob \otimes G$-measure, we just have to invoke Theorem \ref{thm:
measurable maximum} to conclude that $\rho$ is predictable and we
are done.

If $\{ \I \cap \check{\K} = \emptyset \} \cap \Lambda$ does not have
full $\prob \otimes G$-measure, we still have to worry about the
predictable set $\{ \I \cap \check{\K} = \emptyset \} \cap
(\dbra{0,T} \setminus \Lambda)$. On the last set, we consider an
approximating sequence $(\nu_n)_{n \in \Natural}$, keeping every
$\nu_n$ predictable (this is easy to do, since we can choose all
densities $f_n$ to be deterministic --- remember our concrete
example $f_n(x) = \indic_{\{ |x| \leq 1\}} + |x|^{-1/n}
\indic_{\{|x| > 1\}}$); we get a sequence of processes $(\rho_n)_{n
\in \Natural}$ defined on the whole $\dbra{0, T}$ that take values
in $\K \cap \N^\bot$ and solve the corresponding approximating
problems on $\{ \I \cap \check{\K} = \emptyset \} \cap (\dbra{0,T}
\setminus \Lambda)$. According to Lemma \ref{thm:
deterministic_solution}, $(\rho_n)_{n \in \Natural}$ will converge
pointwise to a process $\rho$; this will be predictable (as a
pointwise limit of predictable processes) and satisfy $\,\rel(\pi
\such \rho) \leq 0\,$, $ \, \forall~ \pi \in \Pi_\K$.

Now that we have our candidate $\rho$ for   \num portfolio, we only
need to check its $X$-integrability; according to Lemma \ref{lem:
pred charact of integr up to infty, special} this is   covered by
the criterion $\,\pare{\phi^\rho \cdot G}_t < + \infty\,$ for all $t
\in \dbra{0, T}$. In light of Lemma \ref{lem: necess and suff for
numeraire}, we are done. \qed

\section{On Rates of Convergence to Zero for Positive Supermartingales} \label{sec: Rates of Conv to Zero of Pos Supermarts}

Every positive supermartingale converges as time tends to infinity.
The following decides whether this limit is zero or not in terms of
predictable characteristics, and estimates the rate of convergence
to zero when this is the case.

\begin{prop} \label{prop: rate of supermart convergence }
Let $Z$ be a local supermartingale with $\Delta Z > -1$ and
Doob-Meyer decomposition $Z =  M- A $, where $A$ is an increasing,
predictable process. With $\hat{C} := [Z^\co, Z^\co]$ being  the
quadratic covariation of the continuous local martingale part of $Z$
and $\hat{\eta}$ the predictable compensator of the jump measure
$\hat{\mu}$, define the increasing  predictable process $\,H := A +
\hat{C}/2 + q(1+x)*\hat{\eta}\,$, where $q :\Real_+ \mapsto \Real_+$
is the convex function $q(y) := \bra{-\log a + (1-a^{-1}) y}
\indic_{[0, a)} (y) + \bra{y - 1 - \log y} \indic_{[a,+\infty)}(y)$
for some $a \in (0,1)$.

Consider also the positive supermartingale $Y = \Exp(Z)$. Then, on
the event $\{ H_\infty < + \infty \}$ we have $\lim_{t \to \infty}
Y_t \, \in \, (0, +\infty)\,$, while on $\{ H_\infty = + \infty \}$,
we have $\limsup_{t \to \infty} \left( H^{-1}_t \log Y_t \right) \,
\leq
  \,-1$.
\end{prop}

\medskip

Proposition \ref{prop: rate of supermart convergence } is an
abstract version of Proposition \ref{prop: asymptotic growth
optimality of num}; to obtain that latter proposition from the
former, notice that $W^\pi / W^\rho$ is a positive supermartingale,
and identify the elements $\, A, \hat{C}$ and $q(1+x)
* \hat{\eta}\,$ of Proposition \ref{prop: rate of supermart
convergence } with $\rel(\pi \such \rho) \cdot G$, $(\pi -
\rho)^\top c (\pi - \rho) \cdot G$ and $\pare{\int q_a
\pare{\frac{1+ \pi^\top x}{1+ \rho^\top x}} \nu( \ud x)} \cdot G$.

If we further assume   $\,\Delta Z \geq -1 + \delta\,$ for some
$\delta > 0$, then by considering   $q(x) = x - \log(1+x)$ in the
definition of $H$ we obtain $\,\lim_{t \to \infty}( H^{-1}_t \log
Y_t )= -1\,$  on the set $\{ H_\infty = + \infty \}$;  i.e., we get
the exact rate of decay of $\log Y$ to $- \infty\,$.

\begin{rem} \label{rem: limit theorem for local marts.}
In the course of the proof, we shall make heavy use of the
following: \emph{For a locally square integrable martingale $N$ with
angle-bracket (predictable quadratic variation) process
$\inner{N}{N}$, on the event $\{ \inner{N}{N}_\infty < +\infty \}$
the limit $N_\infty$ exists and is finite, whereas on the event
$\{\inner{N}{N}_\infty = +\infty\}$ we have $\lim_{t \to \infty} N_t
/ \inner{N}{N}_t = 0$}.

Note also that if $N = v(x) * (\hat{\mu} - \hat{\eta})$, then
$\inner{N}{N} \leq v(x)^2 * \hat{\eta}$ (equality holds if and only
if $N$ is quasi-left-continuous). Combining this with the previous
remarks we get that on the event $\{(v(x)^2 * \hat{\eta})_\infty <
+\infty\}$ the limit $N_\infty$ exists and is finite, whereas on
$\{(v(x)^2 * \hat{\eta})_\infty = +\infty\}$ we have $\lim_{t \to
\infty} N_t / (v(x)^2 * \hat{\eta})_t = 0$.
\end{rem}

\proof For the supermartingale $Y = \Exp(Z)$, the stochastic exponential formula (\ref{eq: stoch
expo}) gives $\log Y = Z - [Z^\co, Z^\co]/2 - \sum_{s \leq \cdot} \bra{\Delta Z_s - \log(1 +
\Delta Z_s)}$, or equivalently
\begin{equation}
\log Y = -A + (M^\co - \hat{C}/2) + \big( x*(\hat{\mu} - \hat{\eta})
- [x - \log (1+x)] * \hat{\mu} \big).
\end{equation}

We start with the continuous local martingale part, and use Remark
\ref{rem: limit theorem for local marts.} twice: first, on $\{
\hat{C}_\infty < + \infty \}$, $M^\co_\infty$ exists and is
real-valued; secondly, on $\{ \hat{C}_\infty = + \infty \}$ we get
$\lim_{t \to \infty} (M_t^\co - \hat{C}_t/2)/(\hat{C}_t/2) = -1$.

To deal with the purely discontinuous local martingale part, we
first define the two indicator functions $l := \indic_{[-1,-1+a)}$
and $r := \indic_{[-1+a, +\infty)}$, where $l$ and $r$ stand as
mnemonics for \emph{l}eft and \emph{r}ight. Define the two
semimartingales
\begin{eqnarray*} \label{eq: pur disc broken}
  E &:=& [l (x) \log (1+x)]*\hat{\mu} - [
  l (x) x]*\hat{\eta},
 \\
  F &:=& [r (x) \log (1 + x) ]*(\hat{\mu} - \hat{\eta}) + [r (x)
  q(1+x)]*\hat{\eta}.
\end{eqnarray*}
and observe that $x*(\hat{\mu} - \hat{\eta}) - [x - \log(1 + x)] *
\hat{\mu} = E + F$.

We claim that on $\{\pare{q(1+x)*\hat{\eta}}_\infty < + \infty\}$,
both $E_\infty$ and $F_\infty$ exist and are real-valued. For $E$,
this happens because $\pare{[l(x) q(1+x)]*\hat{\eta}}_\infty < +
\infty$ implies that there will only be a finite number of times
when $\Delta Z \in (-1, -1+a]$ so that both terms in the definition
of $E$ have a limit at infinity. Turning to $F$, the second term in
its definition is obviously finite-valued at infinity whereas for
the local martingale term $[r (x) \log (1 + x) ]*(\hat{\mu} -
\hat{\eta})$ we need only use the set inclusion $\{
\pare{[r(x) q(1+x)]*\hat{\eta}}_\infty < + \infty \} \subseteq \{
\pare{[r (x) \log^2(1 + x)]*\hat{\eta}}_\infty < + \infty \}$ to get that it has finite
predictable quadratic variation and use Remark \ref{rem: limit
theorem for local marts.}.

Now we turn attention to the event
$\{\pare{q(1+x)*\hat{\eta}}_\infty = + \infty\}$; there, at least
one of the quantities $\pare{[l(x) q(1+x)]*\hat{\eta}}_\infty$ and
$\pare{[r (x) q(1+x)]*\hat{\eta}}_\infty$ must be infinite.

On the event $\{ \pare{[r (x) q(1+x)]*\hat{\eta}}_\infty = \infty
\}$, use of the definition of $F$; then Remark \ref{rem: limit
theorem for local marts.} gives $\lim_{t \to \infty} F_t /
\pare{[r(x) q(1+x)]*\hat{\eta}}_t = -1$.

Now let us work on the event $\{ \pare{[l (x) q(1+x)]*\hat{\eta}}_\infty = \infty \}$. We know
that the inequality $\log y \leq y - 1 - q(y)$ holds for $y > 0$; using this last inequality in
the first term in the definition of $E$ we get $E \leq [l (x) (x - q(1+x))]* \hat{\mu} - [l (x)
x]*\hat{\eta}$, or further that $E \leq [l (x) (x -q(1+x))]* (\hat{\mu} - \hat{\eta}) - [l (x)
q(1+x)] * \hat{\eta}$. From this last inequality and Remark \ref{rem: limit theorem for local
marts.} we get $\limsup_{t \to \infty} E_t /
\pare{[l(x) q(1+x)]*\hat{\eta}}_t \leq -1$.

Let us summarize the last paragraphs on the purely discontinuous part. On the event $\{
\pare{q(1+x)*\hat{\eta}}_\infty < + \infty \}$, the limit $\pare{x*(\hat{\mu} - \hat{\eta}) - [x -
\log(1 + x)]
* \hat{\mu}}_\infty$ exists and is finite; on the other hand, on the event
$\{ \pare{q(1+x)*\hat{\eta}}_\infty = + \infty \}$, we have $\limsup_{t \to \infty} \big(
x*(\hat{\mu} - \hat{\eta}) - [x - \log(1 + x)] * \hat{\mu} \big)_t / \big( q(1+x)*\hat{\eta}
\big)_t \leq -1$.

\smallskip

From the previous discussion on the continuous and the purely
discontinuous local martingale parts of $\log Y$ and the definition
of $H$, the result follows. \qed

\section{Proof of Proposition \ref{prop: with num you can never go wrong}} \label{sec: proof that with num you can never go wrong}

\subsection{The proof} \label{subsec: proof that with num you cannot go wrong} Start by
defining $\Omega_0 := \set{(\psi^{\rho} \cdot G)_T < \infty}$ and
$\Omega_A := \Omega \setminus \Omega_0$.

\smallskip

First, we show the result for $\Omega_0$. Assume $\prob[\Omega_0] >
0$, and call $\prob_0$ the probability measure one gets by
conditioning $\prob$ on the set $\Omega_0$. The process $\rho$ of
course remains predictable when viewed under the new measure; and
because we are restricting ourselves on $\Omega_0$, $\rho$ is
$X$-integrable up to $T$ under $\prob_0$.

By a use of the dominated convergence theorem for Lebesgue
and  for stochastic integrals,   all three sequences of processes
$\rho_n \cdot X$, $[\rho_n \cdot X^\co, \rho_n \cdot X^\co]$ and
$\sum_{s \leq \cdot} \bra{\rho_n^\top \Delta X_s - \log (1 +
\rho_n^\top \Delta X_s)}$ converge uniformly (in $t \in [0, T]$) in
$\prob_0$-measure to three processes, that do not depend  on the
sequence $(\rho_n)_{n \in \Natural}$. Then, the stochastic
exponential formula (\ref{eq: stoch expo}) gives that $W^{\rho_n}_T$
converges in $\prob_0$-measure to a random variable, which does not
depend  on the sequence $(\rho_n)_{n \in \Natural}$. Since the limit
of the sequence $(\indic_{\Omega_0} W^{\rho_n}_T)_{n \in \Natural}$
is the same under both the $\prob$-measure and the
$\prob_0$-measure, we conclude that, on $\Omega_0$, the sequence
$(W^{\rho_n}_T)_{n \in \Natural}$ converges in $\prob$-measure to a
real-valued random variable, independently of the choice of the
sequence $(\rho_n)_{n \in \Natural}$.

\smallskip

Now we have to tackle the set $\Omega_A$, which is trickier. We
shall use a ``helping sequence of portfolios''. Suppose
$\prob[\Omega_A] > 0$, otherwise there is nothing to prove. Under
this assumption, there exist a sequence of $[0,1]$-valued
predictable processes $(h_n)_{n \in \Natural}$, such that each
$\pi_n := h_n \rho$ is $X$-integrable up to $T$ and the sequence of
terminal values $((\pi_n \cdot X)_T)_{n \in \Natural}$ is unbounded
in probability (readers unfamiliar with this fact should consult
\cite{Bichteler: stoch. integration}, Corollary 3.6.10, page 128).
It is reasonable to believe (but wrong in general, and a little
tedious to show in our case) that unboundedness in probability of
the terminal values $((\pi_n \cdot X)_T)_{n \in \Natural}$ implies
that the sequence of the terminal values for the \emph{stochastic}
exponentials $(W^{\pi_n}_T)_{n \in \Natural}$ is also unbounded in
probability. We shall show this in Lemma \ref{lem: easy to believe,
but so confusing!} of the next subsection; for the time being, we
accept this as fact. Then $\prob[\limsup_{n \to \infty} W^{\pi_n}_T
= + \infty] > 0$, where the $\limsup$ is taken in probability and
not almost surely (recall Definition \ref{dfn: limsup in prob}).

\smallskip

Let us return to our original sequence of portfolios $(\rho_n)_{n
\in \Natural}$ with $\rho_n = \theta_n \rho$ and show that
$\set{\limsup_{n \to \infty} W^{\pi_n}_T = +\infty} \subseteq
\set{\limsup_{n \to \infty} W^{\rho_n}_T = +\infty}$. Both of
these upper limits, and in fact all the $\limsup$ that will appear
until the end of the proof, are supposed to be in $\prob$-measure.
Since each $\theta_n$ is $[0,1]$-valued and $\lim_{n \to \infty}
\theta_n = \indic$, one can choose an increasing sequence
$\pare{k(n)}_{n \in \Natural}$ of natural numbers such that the
sequence $\big(W^{\theta_{k(n)} \pi_n}_T\big)_{n \in \Natural}$ is
unbounded in $\prob$-measure on the set $\set{\limsup_{n \to
\infty} W^{\pi_n}_T = +\infty}$. Now, each process
$W^{\theta_{k(n)} \pi_n} / W^{\rho_{k(n)}}$ is a positive
supermartingale, since $\rel(\theta_{k(n)} \pi_n \such
\rho_{k(n)}) = \rel(\theta_{k(n)} h_n \rho \such h_n \rho) \leq
0$, the last inequality due to the fact that $[0,1] \ni u \mapsto
\g(u \rho)$ is increasing, and so the sequence of random variables
$\big(W^{\theta_{k(n)} \pi_n}_T / W^{\rho_{k(n)}}_T\big)_{n \in
\Natural}$ is bounded in probability. From the last two facts
follows that the sequence of random variables
$(W^{\rho_{k(n)}}_T)_{n \in \Natural}$ is also unbounded in
$\prob$-measure on $\set{\limsup_{n \to \infty} W^{\pi_n}_T =
+\infty}$.

\smallskip
Up to now we have shown that $\prob[\limsup_{n \to \infty} W^{\rho_n}_T =
+\infty] > 0$, and we also know that $\{ \limsup_{n \to \infty}
W^{\rho_n}_T = +\infty \} \subseteq \Omega_A$; it remains to show
that the last set inclusion is actually an equality (mod $\prob$).
Set $\Omega_B := \Omega_A \setminus \{ \limsup_{n \to \infty}
W^{\rho_n}_T = +\infty \}$ and assume that $\prob[\Omega_B]
> 0$. Working under the conditional measure on $\Omega_B$ (denoted by
$\prob_B$), and following the exact same steps we carried out two
paragraphs ago, we find predictable processes $(h_n)_{n \in
\Natural}$ such that each $\pi_n := h_n \rho$ is $X$-integrable up
to $T$ under $\prob_B$ and such that the sequence of terminal values
$((\pi_n \cdot X)_\infty)_{n \in \Natural}$ is unbounded in
$\prob_B$-probability; then $\prob_B [\limsup_{n \to \infty}
W^{\rho_n}_T = +\infty] > 0$, which contradicts the definition of
$\Omega_B$ and we are done. \qed

\subsection{Unboundedness for Stochastic Exponentials}

We still owe one thing in the previous proof: at some point we had a
sequence of random variables $\pare{(\pi_n \cdot X)_T}_{n \in
\Natural}$   that was unbounded in probability, and  needed to show
that the sequence $(\Exp(\pi_n \cdot X)_T)_{n \in \Natural}$ is
unbounded in probability as well. One has to be careful with
statements like that because, as we shall see in Remark \ref{rem:
stoch expo is not increasing}, the stochastic --- unlike the usual
---  exponential is \emph{not} a monotone operation.

We have to prove the following Lemma \ref{lem: easy to believe, but
so confusing!} and finish the proof of Proposition \ref{prop: with
num you can never go wrong}. To begin, observe that with $R_n :=
\pi_n \cdot X$, the collection $(R_n)_{n \in \Natural}$ is such that
$\Delta R_n > -1$ and $\Exp(R_n)^{-1}$ is a positive supermartingale
for all $n \in \Natural$.

A class $\R$ of semimartingales will be called ``unbounded in
probability'', if the collection $\{ \sup_{t \in [0, T]} \abs{R_t}
\such R \in \R \}$ is unbounded in probability. Similar definitions
apply for (un)boundedness from above and below, taking one-sided
suprema.

\begin{lem} \label{lem: easy to believe, but so confusing!}
Let $\R$ be a collection of semimartingales such that $R_0 = 0$,
$\Delta R > -1$ and $\Exp(R)^{-1}$ is a (positive) supermartingale
for all $R \in \R$ (in particular, $\Exp(R)_T$ exists and takes
values in $(0,\infty]$). Then, the collection of processes $\R$ is
unbounded in probability, if and only if the collection of positive
random variables $\set{\Exp(R)_T \such R \in \R}$ is unbounded in
probability.
\end{lem}

\proof We shall only consider boundedness notions ``in probability''
throughout. Since $R \geq \log \Exp(R)$ for all $R \in \R$, one side
of the equivalence is trivial, and we only have to prove that if
$\R$ is unbounded then $\set{\Exp(R)_T \such R \in \R}$ is
unbounded. We split  the proof of this into four steps.

\smallskip

\noindent (i) Since $\{ \Exp(R)^{-1} \such R \in \R \}$ is a
collection of positive supermartingales, it is bounded from above,
thus $\{ \log \Exp(R) \such R \in \R \}$ is bounded from below.
Since $R \geq \log \Exp(R)$ for all $R \in \R$ and $\R$ is
unbounded, it follows that it \emph{must} be unbounded from above.

\smallskip

\noindent (ii) Let us now show that \emph{the collection of
\emph{random variables} $\{ \Exp(R)_T \such R \in \R \}$ is
unbounded if and only if the collection of \emph{semimartingales}
$\{ \Exp(R) \such R \in \R \}$ is unbounded} (from above, of course,
since they are positive). One direction is trivial: if the
semimartingale class is unbounded, the random variable class is
unbounded too; we only need to argue the reverse implication.
Unboundedness of $\set{\Exp(R) \such R \in \R}$ means that we can
pick an $\epsilon > 0$ so that, for any $n \in \Natural$, there
exists a semimartingale $R^n \in \R$ such that for the stopping
times $\tau_n := \inf \set{t \in [0, T] \such \Exp(R^n)_t \geq n}$
(as usual, we set $\tau_n = \infty$ where the last set is empty) we
have $\prob[\tau_n < \infty] \geq \epsilon$. Each $\Exp(R^n)^{-1}$
is a supermartingale, therefore
\[
\prob[\Exp(R^n)_T^{-1} \leq n^{-1/2}] \geq \prob[\Exp(R^n)_T^{-1}
\leq n^{-1/2} \such \tau_n < \infty] \ \prob[\tau_n < \infty] \geq
\epsilon (1 - n^{-1/2})\,,
\]
so $(\Exp(R^n)_T)_{n \in \Natural}$ is unbounded and the claim of
this paragraph is proved.

\smallskip

We want to show now  that, if $\R$ is unbounded, then $\set{\Exp(R)
\such R \in \R}$ is unbounded too. Define the class $\Z := \set{\Log
\pare{\Exp(R)^{-1}} \such R \in \R}$; we have $Z_0 = 0$, $\Delta Z >
-1$ and that $Z$ is a local supermartingale for all $Z \in \Z$.

\smallskip

\noindent (iii) Let us prove that \emph{if the collection $\Z$ is
bounded from below, then it is also bounded from above}. To this
end, pick any $\epsilon > 0$. We can find an $M \in \Real_+$ such
that the stopping times $\tau_Z := \inf \set{t \in [0, T] \such Z_t
\leq -M + 1}$ (we set $\tau_Z = \infty$ where the last set is empty)
satisfy $\prob[\tau_Z < \infty] \leq \epsilon/2$ for all $Z \in \Z$.
Since $\Delta Z > -1$, we have $Z_{\tau_Z} \geq -M$ and so each
stopped process $Z^{\tau_Z}$ is a supermartingale (it is a local
supermartingale bounded uniformly from below). Then, with
$y_\epsilon := 2M/\epsilon$ we have
\[
\prob \Big[ \sup_{t \in [0, T]} Z_t > y_\epsilon \Big] \leq
(\epsilon/2) + \prob \Big[ \sup_{t \in [0, T]} Z^{\tau_Z}_t >
y_\epsilon \Big] \leq (\epsilon/2) + (1 + y_\epsilon/M)^{-1} \leq
\epsilon\,,
\]
and thus $\Z$ is bounded from above too.

\smallskip

\noindent (iv) Now we have all the ingredients for the proof.
Suppose that $\R$ is unbounded; we have seen that it has to be
unbounded from above. Using Lemma \ref{lem: inverse of stochastic
expo} with $Y \equiv 0$, we get that every $Z \in \Z$ is of the form
\begin{equation} \label{eq: inverse of stoch expo special}
Z  = - R + [R^\co,R^\co] + \sum_{s \leq \cdot} \frac{|\Delta
R_s|^2}{1 + \Delta R_s}.
\end{equation}
When $\Z$ is unbounded from below, things are pretty simple, because
$\log \Exp(Z) \leq Z$ for all $Z \in \Z$ so that $\set{\log \Exp(\Z)
\such Z \in \Z}$ is unbounded from below and thus $\set{\Exp(R)
\such R \in \R} = \set{\exp (- \log\Exp(Z)) \such Z \in \Z}$ is
unbounded from above.

It remains to see what happens if $\Z$ is bounded from below. From
step (iii) we know that $\Z$ must be bounded from above as well.
Then, because of equation (\ref{eq: inverse of stoch expo}) and the
unboundedness from above of $\R$, the
collection $\{ \ [R^\co,R^\co] + \sum_{s \leq \cdot} \bra{|\Delta
R_s|^2/\pare{1 + \Delta R_s}} \such R \in \R \ \}$ of increasing
processes is also unbounded. Now, for $Z \in \Z$ we have
\[
\log \Exp(Z) = - \log \Exp(R) = - R + \frac{1}{2} [R^\co, R^\co] +
\sum_{s \leq \cdot} \bra{\Delta R_s - \log(1 + \Delta R_s)}
\]
from (\ref{eq: inverse of stoch expo special}) and the stochastic
exponential formula, so that
\[
Z - \log \Exp (Z) = \frac{1}{2} [R^\co, R^\co] + \sum_{s \leq \cdot} \bra{\log(1 + \Delta R_s) -
\frac{\Delta R_s}{1 + \Delta R_s}}.
\]
The collection of increasing processes on the right-hand-side of
this last equation is unbounded, because $\{ \ [R^\co,R^\co] +
\sum_{s \leq \cdot} \bra{(\Delta R_s)^2/\pare{1 + \Delta R_s}} \such
R \in \R \ \}$ is unbounded too, as we observed. But since $\Z$ is
bounded, this means that $\set{\log \Exp (\Z) \such Z \in \Z}$ is
unbounded from below, and we conclude again as before. \qed

\begin{rem} \label{rem: stoch expo is not increasing}
Without the assumption that $\{ \Exp(R)^{-1} \such R \in \R \}$
consists of supermartingales, this result is no longer true. In
fact, take $T \equiv + \infty$ and $\R = \{ R \}$ where $R_t = a t +
\beta_t$, with $a \in (0,1/2)$ and $\beta$ is a standard
1-dimensional Brownian motion. Then, $R$ is bounded from below and
unbounded from above, nevertheless $\log \Exp (R)_t = (a - 1/2) t +
\beta_t$ is bounded from above, and unbounded from below.
\end{rem}

\appendix

\section{Measurable Random Subsets} \label{subsec:
measurable selection}

Throughout this section we shall be working on a measurable space
$(\Omegat, \Pre)$; although the results are general, think of
$\Omegat$ as $\Omega \times \Real_+$ and of $\Pre$ as the
\emph{predictable} $\sigma$-algebra. The metric of the Euclidean
space $\Real^d$, its denoted by ``$\dist$'' and its generic point by
$z$. Proofs of the results below will not be given, but  can be
found (in greater generality) in Chapter 17 of \cite{Aliprantis};
for shorter proofs of the specific results, see \cite{K: thesis}.
The subject of measurable random subsets and measurable selection is
slightly gory in its technicalities, but the statements should be
intuitively clear.

A \textsl{random subset of $\Real^d$} is just a random variable
taking values in $2^{\Real^d}$, the powerset (class of all subsets)
of $\Real^d$. Thus, a random subset of $\Real^d$ is a function $\Af
: \Omegat \mapsto 2^{\Real^d}$. A random subset $\Af$ of $\Real^d$
will be called \textsl{closed} (resp., \textsl{convex}) if the set
$\Af(\omegat)$ is closed (resp., convex) for \emph{every} $\omegat
\in \Omegat$.

Measurability requirements on random subsets are given by placing
some measurable structure on the space $2^{\Real^d}$, which we endow
with the smallest $\sigma$-algebra that makes   the mappings
$2^{\Real^d} \ni A \mapsto \dist(z, A) \in \Real_+ \cup \{ + \infty
\}$ measurable for all $z \in \Real^d$ (by definition, $\dist(z,
\emptyset) = +\infty$). The following equivalent formulations are
sometimes useful.

\begin{prop} \label{prop: sigma-algebra on powerset equal to others}
The constructed $\sigma$-algebra on $2^{\Real^d}$ is also the
smallest $\sigma$-algebra that makes the class $\{ 2^{\Real^d} \in A
\mapsto \indic_{\{ A \cap K \neq \emptyset \}} \}$, for every
compact (resp. closed, resp. open) $K \subseteq \Real^d\,$ of
functions measurable.
\end{prop}

From Proposition \ref{prop: sigma-algebra on powerset equal to
others}, a random subset $\Af$ of $\Real^d$ is measurable if for any
compact $K \subseteq \Real^d$, the set $ \{\Af \cap K \neq \emptyset
\} := \{ \omegat \in \Omegat \such \Af(\omegat) \cap K \neq
\emptyset \}$ is $\Pre$-measurable.

\begin{rem} \label{rem: measurable singleton is classical
measurable} Suppose that the random subset $\Af$ is a
\emph{singleton} $\Af(\omegat) = \{ a(\omegat) \}$ for some $a :
\Omegat \mapsto \Real^d$. Then, $\Af$ is measurable if and only if
$\{ a \in K \} \in \Pre$ for all closed $K \subseteq \Real^d$, i.e.,
if and only if $a$ is $\Pre$-measurable.
\end{rem}

We now deal with unions and intersections of random subsets of
$\Real^d$.

\begin{lem} \label{lem: union and intersects of measur set-valued processes}
Suppose that $\pare{\Af_n}_{n \in \Natural}$ is a sequence of
measurable random subsets of $\Real^d$. Then, the union
$\,\bigcup_{n \in \Natural} \Af_n$ is also measurable. If,
furthermore, each random subset $\,\Af_n$ is closed, then the
intersection $\,\bigcap_{n \in \Natural} \Af_n$ is measurable.
\end{lem}

The following lemma gives a way to construct measurable, closed
random subsets of $\Real^d$. To state it, we shall need (a slight
generalization of) the notion of   Carath\'eodory function. For a
measurable closed random subset $\Af$ of $\Real^d$, a mapping $f$ of
$\Omegat \times \Real^d$ into another topological space will be
called \textsl{Carath\'eodory on $\Af$}, if it is measurable (with
respect to the product $\sigma$-algebra on $\Omegat \times
\Real^d$), and if $\, z \mapsto f(\omegat, z)$ is continuous on
$\Af(\omegat)$, for each $\omegat \in \Omegat$. Of course, if $\,\Af
\equiv \Real^d$, we recover the usual textbook notion of a
Carath\'eodory function.

\begin{lem} \label{lem: Caratheodory makes measurable sets}
Let $E$ be any topological space, $F \subseteq E$ a closed subset,
and $\Af$ a closed and convex random subset of $\Real^d$. If $f :
\Omegat \times \Real^d \to E$ is a Carath\'eodory function on
$\Af$, then $\K := \{ z \in \Af \such f(\cdot, z) \in F \}$ is
closed and measurable.
\end{lem}

The last result focuses on the measurability of the ``argument''
process in random optimization problems.

\begin{thm} \label{thm: measurable maximum}
Suppose that $\Af$ is a closed and convex, measurable, non-empty
random subset of $\Real^d$, and $f : \Omegat \times \Real^d \mapsto
\Real \cup \{ - \infty \}$ is a Carath\'eodory function on $\Af$.
For the optimization problem $\,f_* (\omegat) = \sup_{z \in \Af}
f(\omegat, z)\,$, we have:
\begin{enumerate}
    \item The value function $f_*$ is $\Pre$-measurable.
    \item Suppose that  $f_*(\omegat)$ is finite for all $\omegat$,
    and that there exists a unique $z_*(\omegat) \in \Af(\omegat)$
    satisfying $f(\omegat, z_*(\omegat)) = f_* (\omegat)$. Then
    $\, \omegat \mapsto z_* (\omegat)\,$ is $\,\Pre$-measurable.
\end{enumerate}
\noindent
 In particular, if $\Af$ is a closed and convex,
measurable, non-empty random subset of $\Real^d$, we can find a
$\Pre$-measurable $h : \Omegat \to \Real^d$ with $h(\omegat) \in
\Af(\omegat)$ for all $\omegat \in \Omegat$.
\end{thm}

For the ``particular'' case of the last theorem one can use for
example the function $f(x) = - |x|$ and the result first part of the
theorem.

\section{Semimartingales and Stochastic
Integration up to $+\infty$}\label{subsec: semimarts up to infty}

\noindent We recall here a few important concepts from \cite{Cherny}
and prove a few useful results. One can also check \cite{DM-2} for
the ideas presented below.

\begin{defn} \label{dfn: semimart up to inf}
Let $X = (X_t)_{t \in \Real_+}$ be a semimartingale such that
$X_\infty := \lim_{t \to \infty} X_t$ exists. Then $X$ will be
called a \textsl{semimartingale up to infinity} if the process
$\tilde{X}$ defined on the time interval $[0,1]$ by $\tilde{X} (t) =
X \big( t / (1-t) \big)$ (of course, $\tilde{X}_1 = X_\infty$) is a
semimartingale relative to the filtration $\tilde{\mathbf{F}} =
(\tilde{\F}_t)_{t \in [0,1]}$ defined by $\tilde{\F}_t := \F_{ t /
(1-t)}$ for $0 \leq t < 1$ and $\tilde{\F}_1 := \bigvee_{t \in
\Real_+} \F_t$.

Similarly, we define local martingales up to infinity, processes of
finite variation up to infinity, etc., if the corresponding process
$\tilde{X}$ has the property.

Fix a $d$-dimensional semimartingale $X$. An $X$-integrable
predictable process $\pi$ will be called \textsl{$X$-integrable up
to infinity} if $\pi \cdot X$ is a semimartingale up to infinity.

\end{defn}

To appreciate the difference between a semimartingale with limit at
infinity and a semimartingale up to infinity, consider the simple
example where $X$ is the deterministic, continuous process $X_t :=
t^{-1} \sin t$; then $X$ is a semimartingale with $X_\infty = 0$,
but $\Var(X)_\infty = + \infty$ and thus $X$ cannot be a
semimartingale up to infinity (a deterministic semimartingale must
be of finite variation).

Every semimartingale up to infinity $X$ can be written as the sum $X
= A + M$, where $A$ is a process of finite variation up to infinity
(which simply means that $\Var(A)_\infty < \infty$) and $M$ is a
local martingale up to infinity (which means that there exists an
increasing sequence of stopping times $(T_n)_{n \in \Natural}$ with
$\set{T_n = + \infty} \uparrow \Omega$ such that each of the stopped
processes $M^{T_n}$ is a uniformly integrable martingale).

\begin{lem} \label{lem: positive supermart is semimart to infty}
A positive supermartingale $Z$ is a special semimartingale up to
infinity. If furthermore $Z_\infty > 0$, then $\Log(Z)$ is also a
special semimartingale up to infinity, and both $Z^{-1}$ and
$\Log(Z^{-1})$ are semimartingales up to infinity.
\end{lem}

\proof We start with the Doob-Meyer decomposition $Z = M - A$, where
$M$ is a local martingale with $M_0 = Z_0$ and $A$ is an increasing,
predictable process. The positive local martingale $M$ is a
supermartingale, and we can infer that both limits $Z_\infty$ and
$M_\infty$ exist and are integrable. This means that $A_\infty$
exists and actually $\expec[A_\infty] = \expec[M_\infty] -
\expec[Z_\infty] < \infty$, so $A$ is a predictable process of
integrable variation up to infinity. It remains to show that $M$ is
a local martingale up to infinity. Set $T_n := \inf \set{t \geq 0
\such M_t \geq n}$; this obviously satisfies $\set{T_n = + \infty}
\uparrow \Omega$ (the supremum of a positive supermartingale is
finite). Since $\sup_{0 \leq t \leq T_n} M_t \leq n + M_{T_n}
\indic_{\set{T_n < \infty}}$ and by the optional sampling theorem
$\expec[M_{T_n} \indic_{\set{T_n < \infty}}] \leq \expec[M_0] <
\infty$, we get $\expec[\sup_{0 \leq t \leq T_n} M_t] < \infty$.
Thus, the local martingale $M^{T_n}$ is actually a uniformly
integrable martingale and thus $Z$ is a special semimartingale up to
infinity.

Now assume that $Z_\infty > 0$. Since $Z$ is a supermartingale, this
will mean that both $\tilde{Z}$ and $\tilde{Z}_-$ are bounded away
from zero. (A ``tilde'' over a process means that we are considering
the process of Definition \ref{dfn: semimart up to inf} under the
new filtration $\tilde{\mathbf{F}}$.) Since $\tilde{Z}^{-1}_-$ is
locally bounded and $\tilde{Z}$ is a special semimartingale,
$\Log(\tilde{Z}) = \tilde{Z}_-^{-1} \cdot \tilde{Z}$ will be a
special semimartingale as well, meaning that $\Log(Z)$ is a special
semimartingale up to infinity. Furthermore, It\^o's formula applied
to the inverse function $(0,\infty) \ni x \mapsto x^{-1}$ implies
that $\tilde{Z}^{-1}$ is a semimartingale up to infinity and since
$\tilde{Z}_-$ is locally bounded, $\Log(\tilde{Z}^{-1}) =
\tilde{Z}_- \cdot \tilde{Z}^{-1}$ is a semimartingale, which
finishes the proof. \qed

\begin{rem} \label{rem: semimarts up to T from semimart up to inf}
In this paper we consider ``semimartingales up to time $T$'' and
``stochastic integration up to time $T$'' where $T$ is a stopping
time rather than ``semimartingales up to infinity'' and ``stochastic
integration up to infinity''. One can use all the results of this
section applying them to the processes stopped at time $T$ ---
divergence from the usual notion of integrability appears only when
$\prob[T = \infty] > 0$.
\end{rem}

\section{$\sigma$-Localization} \label{subsec: sigma localization}

A good account of the concept of $\sigma$-localization is given in
Kallsen \cite{Kallsen: sigma-localization}. Here we recall briefly
what is needed for our purposes. For a semimartingale $Z$ and a
predictable set $\Sigma$, define $Z^\Sigma := \indic_\Sigma \cdot
Z$.

\begin{defn}
Let $\Z$ be a class of semimartingales. Then, the corresponding
$\sigma$-localized class $\Z_\sigma$ is defined as the set of all
semimartingales $Z$ for which there exists an increasing sequence
$(\Sigma_n)_{n \in \Natural}$ of predictable sets, such that
$\Sigma_n \uparrow \Omega \times \Real_+$ (up to   evanescence)
and $Z^{\Sigma_n} \in \Z$ for all $n \in \Natural$.
\end{defn}

When the corresponding class $\Z$ has a name (like
``supermartingales'') we baptize the class $\Z_\sigma$ with the same
name preceded by ``$\sigma$-'' (like ``$\sigma$-supermartingales'').

The concept of $\sigma$-localization is a natural extension of the
well-known concept of localization along a sequence $(\tau_n)_{n \in
\Natural}$ of stopping times, as one can easily see by considering
the predictable sets $\Sigma_n \equiv \dbra{0, \tau_n} := \{(\omega,
t) \in \Omega \times \Real_+ \such 0 \leq t \leq \tau_n(\omega) \}$.

Let us define the set $\Uc$ of semimartingales $Z$, such that the
collection of random variables $\{ Z_\tau \such \tau \text{ is a
stopping time} \}$ is uniformly integrable --- also known in the
literature as semimartingales of class (D). The elements of $\Uc$
admit the \emph{Doob-Meyer decomposition} $Z = A + M$ into a
predictable finite variation part $A$ with $A_0 = 0$ and $\expec
[\Var(A)_\infty] < \infty$ and a uniformly integrable martingale
$M$. It is then obvious that the localized class $\Uc_\loc$
corresponds to all special semimartingales; they are exactly the
ones which admit a Doob-Meyer decomposition as before, but where now
$A$ is only a predictable, finite variation process with $A_0 = 0$
and $M$ a local martingale. Let us remark that the local
supermartingales (resp., local submartingales) correspond to these
elements of $\Uc_\loc$ with $A$ decreasing (resp., increasing). This
last result can be found for example in Jacod's book \cite{Jacod:
calcul stochastic}.

One can have very intuitive interpretation of some
$\sigma$-localized classes in terms of the predictable
characteristics of $Z$.

\begin{prop} \label{prop: pred characterization of (super)marts}
Consider a scalar semimartingale $Z$, and let $(b, c, \nu)$ be the
triplet of predictable characteristics of $Z$ relative to the
canonical truncation function and the operational clock $G$. Then
\begin{enumerate}
    \item $Z$ belongs to $\Uc_\loc$ if and only if the predictable process
    $\int \abs{x} \indic_{\{ \abs{x} > 1 \}} \nu (\ud x)$ is
    $G$-integrable;
    \item $Z$ belongs to $\Uc_\sigma$ if and only if
    $\int \abs{x} \indic_{\{ \abs{x} > 1 \}} \nu (\ud x) <
    \infty$; and
    \item $Z$ is a $\sigma$-supermartingale, if and only if we have 
    $\int \abs{x} \indic_{\{ \abs{x} > 1 \}} \nu (\ud x) < +
    \infty$
    and $b + \int x \indic_{\{ \abs{x} > 1 \}} \nu (\ud x) \leq 0$.
\end{enumerate}
\end{prop}

\proof The first statement follows from the fact that a
1-dimensional semimartingale $Z$ is a special semimartingale (i.e.,
a member of $\Uc_\loc$) if and only if $\bra{\abs{x} \indic_{\{
\abs{x} > 1 \}}} * \hat{\eta}$ is a finite, increasing predictable
process (one can consult Jacod \cite{Jacod: calcul stochastic} for
this fact). The second statement follows easily from the first and
$\sigma$-localization. Finally, the third follows for the fact that
for a process in $\Uc_\loc$ the predictable finite variation part is
given by the process $\pare{b + \int \bra{ x \indic_{\{ \abs{x} > 1
\}}} \nu(\ud x)} \cdot G\,$,   using the last remark before the
proposition, the first part of the proposition, and
$\sigma$-localization. \qed

\medskip

Results like the last proposition are very intuitive, because $b +
\int x \indic_{\{ \abs{x} > 1 \}} \nu (\ud x)$ represents the
infinitesimal drift rate   of the semimartingale $Z$; we expect
this rate to be negative (resp., positive) in the case of
$\sigma$-supermartingales (resp., $\sigma$-submartingales). The
importance of $\sigma$-localization is that it allows us to talk
directly about drift \emph{rates} of processes, rather than about
drifts. Sometimes drift rates exist, but cannot be integrated to
give a drift process; this is when the usual localization
technique fails, and the concept of $\sigma$-localization becomes
useful.

The  following result gives sufficient conditions for  a
$\sigma$-supermartingale to be a local supermartingale (or even
plain supermartingale).

\begin{prop} \label{prop: pos sigma supermat is supermart}
Suppose that $Z$ is a scalar semimartingale with triplet of
predictable characteristics $(b,c,\nu)$.
\begin{enumerate}
    \item Suppose that $Z$ is a
    $\sigma$-supermartingale. Then, the following are equivalent:
    \begin{enumerate}
        \item $Z$ is a local supermartingale.
        \item The positive, predictable process $\int (-x) \indic_{\{ x < -1 \}} \nu (\ud x)$ is $G$-integrable.
    \end{enumerate}
    \item If $Z$ is a $\sigma$-supermartingale (resp., $\sigma$-martingale) and bounded from below by a constant,
    then it is a local supermartingale (resp., local martingale). If furthermore $\expec [Z_0^+] < \infty$, it is a supermartingale.
    \item If $Z$ is bounded from below by a constant, then it is a supermartingale if and only if $\expec [Z_0^+] <
    \infty$ and $b + \int x \indic_{\{ \abs{x} > 1 \}} \nu (\ud x) \leq
    0$.
\end{enumerate}
\end{prop}

\proof For the proof of (1), the implication (a) $\Rightarrow$ (b) follows from part (1) of
Proposition \ref{prop: pred characterization of (super)marts}. For (b) $\Rightarrow$ (a), assume
that $\int (-x) \indic_{\{ x < - 1 \}} \nu (\ud x)$ is $G$-integrable. Since $Z$ is a
$\sigma$-supermartingale, it follows from part (3) of Proposition \ref{prop: pred characterization
of (super)marts} that $\int x \indic_{\{ x > 1 \}} \nu (\ud x) \leq - b + \int (-x) \indic_{\{ x <
- 1 \}} \nu (\ud x)$, and therefore $\int \abs{x} \indic_{\{ \abs{x}
> 1 \}} \nu (\ud x) \leq -b + 2 \int (-x) \indic_{\{ x < - 1 \}} \nu (\ud x)$.
The last dominating process is $G$-integrable, thus $Z \in \Uc_\loc$
(again, part (1) of Proposition \ref{prop: pred characterization of
(super)marts}). The special semimartingale $Z$ has predictable
finite variation part equal to $\pare{b + \int x \indic_{\{ x >  1
\}} \nu (\ud x)} \cdot G$, which is decreasing, so that $Z$ is a
local supermartingale.

For part (2), we can of course assume that $Z$ is positive. We discuss the case of a
$\sigma$-supermartingale; the $\sigma$-martingale case follows in the same way. According to part
(1) of this proposition, we only need to show that $\int (-x) \indic_{\{ x < - 1 \}} \nu (\ud x)$
is $G$-integrable. But since the negative jumps of $Z$ are bounded in magnitude by $Z_-$, we have
that $\int (-x) \indic_{\{ x < - 1 \}} \nu (\ud x) \leq (Z_-) \nu \bra{ x < -1 }$, which is
$G$-integrable, because $\nu \bra{ x < -1}$ is $G$-integrable and $Z_-$ is locally bounded. Now,
if we further assume that $\expec [Z_0] < \infty$, Fatou's lemma for conditional expectations
gives us that the positive local supermartingale $Z$ is a supermartingale.

Let us move on to part (3) and assume that $Z$ is positive. First
assume that $Z$ is a supermartingale. Then, of course we have
$\expec [Z_0] < \infty$ and that $Z$ is an element of $\Uc_\sigma$
(and even of $\Uc_\loc$) and part (3) of Proposition \ref{prop: pred
characterization of (super)marts} ensures that $b + \int x
\indic_{\{ \abs{x} > 1 \}} \nu (\ud x) \leq 0$. Now, assume that $Z$
is a positive semimartingale with $\expec [Z_0] < \infty$ and that
$b + \int x \indic_{\{ \abs{x} > 1 \}} \nu (\ud x) \leq 0$. Then, of
course we have that $\int x \indic_{\{ x
> 1 \}} \nu (\ud x) < \infty$. Also, since $Z$ is positive we always have that $\nu \bra{x < -Z_-} =
0$ so that $\int (-x) \indic_{\{ x < - 1 \}} \nu (\ud x) < \infty$
too. Part (2) of Proposition \ref{prop: pred characterization of
(super)marts} will give us that $Z \in \Uc_\sigma$, and part (3) of
the same proposition that $Z$ is a $\sigma$-supermartingale.
Finally, part (2) of this proposition gives us that $Z$ is a
supermartingale. \qed

\medskip

The special case of result (3) of Proposition \ref{prop: pos sigma
supermat is supermart} when $Z$ is a $\sigma$-martingale is
sometimes called ``The Ansel-Stricker theorem'', since it first
appeared (in a slightly different, but equivalent form) in
\cite{Ansel-Stricker: couverture}. In \cite{Kallsen:
sigma-localization}, one can find the proof of the case when $Z$ is
a $\sigma$-supermartingale bounded from below with $\expec [Z_0^+] <
\infty$.


\end{document}